\documentclass[lettersize,10pt,journal]{IEEEtran}
\IEEEoverridecommandlockouts

\usepackage{cite}
\usepackage[T1]{fontenc}
\usepackage{graphicx}
\usepackage{amssymb}
\usepackage{amsmath}
\usepackage{amsthm}
\usepackage{subfigure}
\usepackage{booktabs}
\usepackage{multirow} 
\usepackage{microtype}
\usepackage{balance}
\usepackage{xcolor}

\newcommand{\red}[1]{\textcolor{red}{#1}}
\definecolor{mygreen}{RGB}{34, 139, 34} 
\newcommand{\green}[1]{\textcolor{mygreen}{#1}}

\usepackage{amssymb}
\usepackage{amsfonts}
\usepackage{mathrsfs}
\usepackage{xspace}
\usepackage{bm}
\usepackage{upgreek}

\newcommand{\safemath}[2]{\newcommand{#1}{\ensuremath{#2}\xspace}}

\safemath{\bma}{\mathbf{a}}
\safemath{\bmb}{\mathbf{b}}
\safemath{\bmc}{\mathbf{c}}
\safemath{\bmd}{\mathbf{d}}
\safemath{\bme}{\mathbf{e}}
\safemath{\bmf}{\mathbf{f}}
\safemath{\bmg}{\mathbf{g}}
\safemath{\bmh}{\mathbf{h}}
\safemath{\bmi}{\mathbf{i}}
\safemath{\bmj}{\mathbf{j}}
\safemath{\bmk}{\mathbf{k}}
\safemath{\bml}{\mathbf{l}}
\safemath{\bmm}{\mathbf{m}}
\safemath{\bmn}{\mathbf{n}}
\safemath{\bmo}{\mathbf{o}}
\safemath{\bmp}{\mathbf{p}}
\safemath{\bmq}{\mathbf{q}}
\safemath{\bmr}{\mathbf{r}}
\safemath{\bms}{\mathbf{s}}
\safemath{\bmt}{\mathbf{t}}
\safemath{\bmu}{\mathbf{u}}
\safemath{\bmv}{\mathbf{v}}
\safemath{\bmw}{\mathbf{w}}
\safemath{\bmx}{\mathbf{x}}
\safemath{\bmy}{\mathbf{y}}
\safemath{\bmz}{\mathbf{z}}
\safemath{\bmzero}{\mathbf{0}}
\safemath{\bmone}{\mathbf{1}}

\bmdefine{\biad}{a}
\bmdefine{\bibd}{b}
\bmdefine{\bicd}{c}
\bmdefine{\bidd}{d}
\bmdefine{\bied}{e}
\bmdefine{\bifd}{f}
\bmdefine{\bigd}{g}
\bmdefine{\bihd}{h}
\bmdefine{\biid}{i}
\bmdefine{\bijd}{j}
\bmdefine{\bikd}{k}
\bmdefine{\bild}{l}
\bmdefine{\bimd}{m}
\bmdefine{\bind}{n}
\bmdefine{\biod}{o}
\bmdefine{\bipd}{p}
\bmdefine{\biqd}{q}
\bmdefine{\bird}{r}
\bmdefine{\bisd}{s}
\bmdefine{\bitd}{t}
\bmdefine{\biud}{u}
\bmdefine{\bivd}{v}
\bmdefine{\biwd}{w}
\bmdefine{\bixd}{x}
\bmdefine{\biyd}{y}
\bmdefine{\bizd}{z}

\bmdefine{\bixid}{\xi}
\bmdefine{\bilambdad}{\lambda}
\bmdefine{\bimud}{\mu}
\bmdefine{\bithetad}{\theta}
\bmdefine{\biphid}{\phi}
\bmdefine{\bideltad}{\delta}

\safemath{\bmia}{\biad}
\safemath{\bmib}{\bibd}
\safemath{\bmic}{\bicd}
\safemath{\bmid}{\bidd}
\safemath{\bmie}{\bied}
\safemath{\bmif}{\bifd}
\safemath{\bmig}{\bigd}
\safemath{\bmih}{\bihd}
\safemath{\bmii}{\biid}
\safemath{\bmij}{\bijd}
\safemath{\bmik}{\bikd}
\safemath{\bmil}{\bild}
\safemath{\bmim}{\bimd}
\safemath{\bmin}{\bind}
\safemath{\bmio}{\biod}
\safemath{\bmip}{\bipd}
\safemath{\bmiq}{\biqd}
\safemath{\bmir}{\bird}
\safemath{\bmis}{\bisd}
\safemath{\bmit}{\bitd}
\safemath{\bmiu}{\biud}
\safemath{\bmiv}{\bivd}
\safemath{\bmiw}{\biwd}
\safemath{\bmix}{\bixd}
\safemath{\bmiy}{\biyd}
\safemath{\bmiz}{\bizd}

\safemath{\bmxi}{\bixid}
\safemath{\bmlambda}{\bilambdad}
\safemath{\bmmu}{\bimud}
\safemath{\bmtheta}{\bithetad}
\safemath{\bmphi}{\biphid}
\safemath{\bmdelta}{\bideltad}

\safemath{\bA}{\mathbf{A}}
\safemath{\bB}{\mathbf{B}}
\safemath{\bC}{\mathbf{C}}
\safemath{\bD}{\mathbf{D}}
\safemath{\bE}{\mathbf{E}}
\safemath{\bF}{\mathbf{F}}
\safemath{\bG}{\mathbf{G}}
\safemath{\bH}{\mathbf{H}}
\safemath{\bI}{\mathbf{I}}
\safemath{\bJ}{\mathbf{J}}
\safemath{\bK}{\mathbf{K}}
\safemath{\bL}{\mathbf{L}}
\safemath{\bM}{\mathbf{M}}
\safemath{\bN}{\mathbf{N}}
\safemath{\bO}{\mathbf{O}}
\safemath{\bP}{\mathbf{P}}
\safemath{\bQ}{\mathbf{Q}}
\safemath{\bR}{\mathbf{R}}
\safemath{\bS}{\mathbf{S}}
\safemath{\bT}{\mathbf{T}}
\safemath{\bU}{\mathbf{U}}
\safemath{\bV}{\mathbf{V}}
\safemath{\bW}{\mathbf{W}}
\safemath{\bX}{\mathbf{X}}
\safemath{\bY}{\mathbf{Y}}
\safemath{\bZ}{\mathbf{Z}}

\safemath{\bZero}{\mathbf{0}}
\safemath{\bOne}{\mathbf{1}}
\safemath{\bDelta}{\mathbf{\Delta}}
\safemath{\bLambda}{\mathbf{\UpLambda}}
\safemath{\bPhi}{\mathbf{\Upphi}}
\safemath{\bSigma}{\mathbf{\Upsigma}}
\safemath{\bOmega}{\mathbf{\Upomega}}
\safemath{\bTheta}{\mathbf{\Uptheta}}

\bmdefine{\biAd}{A}
\bmdefine{\biBd}{B}
\bmdefine{\biCd}{C}
\bmdefine{\biDd}{D}
\bmdefine{\biEd}{E}
\bmdefine{\biFd}{F}
\bmdefine{\biGd}{G}
\bmdefine{\biHd}{H}
\bmdefine{\biId}{I}
\bmdefine{\biJd}{J}
\bmdefine{\biKd}{K}
\bmdefine{\biLd}{L}
\bmdefine{\biMd}{M}
\bmdefine{\biOd}{N}
\bmdefine{\biPd}{O}
\bmdefine{\biQd}{P}
\bmdefine{\biRd}{R}
\bmdefine{\biSd}{S}
\bmdefine{\biTd}{T}
\bmdefine{\biUd}{U}
\bmdefine{\biVd}{V}
\bmdefine{\biWd}{W}
\bmdefine{\biXd}{X}
\bmdefine{\biYd}{Y}
\bmdefine{\biZd}{Z}

\bmdefine{\biDelta}{\Delta}
\bmdefine{\biLambda}{\Lambda}
\bmdefine{\biPhi}{\Phi}
\bmdefine{\biSigma}{\Sigma}
\bmdefine{\biOmega}{\Omega}
\bmdefine{\biTheta}{\Theta}

\safemath{\bimA}{\biAd}
\safemath{\bimB}{\biBd}
\safemath{\bimC}{\biCd}
\safemath{\bimD}{\biDd}
\safemath{\bimE}{\biEd}
\safemath{\bimF}{\biFd}
\safemath{\bimG}{\biGd}
\safemath{\bimH}{\biHd}
\safemath{\bimI}{\biId}
\safemath{\bimJ}{\biJd}
\safemath{\bimK}{\biKd}
\safemath{\bimL}{\biLd}
\safemath{\bimM}{\biMd}
\safemath{\bimN}{\biNd}
\safemath{\bimO}{\biOd}
\safemath{\bimP}{\biPd}
\safemath{\bimQ}{\biQd}
\safemath{\bimR}{\biRd}
\safemath{\bimS}{\biSd}
\safemath{\bimT}{\biTd}
\safemath{\bimU}{\biUd}
\safemath{\bimV}{\biVd}
\safemath{\bimW}{\biWd}
\safemath{\bimX}{\biXd}
\safemath{\bimY}{\biYd}
\safemath{\bimZ}{\biZd}

\safemath{\bimDelta}{\biDelta}
\safemath{\bimLambda}{\biLambda}
\safemath{\bimPhi}{\biPhi}
\safemath{\bimSigma}{\biSigma}
\safemath{\bimOmega}{\biOmega}
\safemath{\bimTheta}{\biTheta}

\safemath{\setA}{\mathcal{A}}
\safemath{\setB}{\mathcal{B}}
\safemath{\setC}{\mathcal{C}}
\safemath{\setD}{\mathcal{D}}
\safemath{\setE}{\mathcal{E}}
\safemath{\setF}{\mathcal{F}}
\safemath{\setG}{\mathcal{G}}
\safemath{\setH}{\mathcal{H}}
\safemath{\setI}{\mathcal{I}}
\safemath{\setJ}{\mathcal{J}}
\safemath{\setK}{\mathcal{K}}
\safemath{\setL}{\mathcal{L}}
\safemath{\setM}{\mathcal{M}}
\safemath{\setN}{\mathcal{N}}
\safemath{\setO}{\mathcal{O}}
\safemath{\setP}{\mathcal{P}}
\safemath{\setQ}{\mathcal{Q}}
\safemath{\setR}{\mathcal{R}}
\safemath{\setS}{\mathcal{S}}
\safemath{\setT}{\mathcal{T}}
\safemath{\setU}{\mathcal{U}}
\safemath{\setV}{\mathcal{V}}
\safemath{\setW}{\mathcal{W}}
\safemath{\setX}{\mathcal{X}}
\safemath{\setY}{\mathcal{Y}}
\safemath{\setZ}{\mathcal{Z}}
\safemath{\emptySet}{\varnothing}

\safemath{\colA}{\mathscr{A}}
\safemath{\colB}{\mathscr{B}}
\safemath{\colC}{\mathscr{C}}
\safemath{\colD}{\mathscr{D}}
\safemath{\colE}{\mathscr{E}}
\safemath{\colF}{\mathscr{F}}
\safemath{\colG}{\mathscr{G}}
\safemath{\colH}{\mathscr{H}}
\safemath{\colI}{\mathscr{I}}
\safemath{\colJ}{\mathscr{J}}
\safemath{\colK}{\mathscr{K}}
\safemath{\colL}{\mathscr{L}}
\safemath{\colM}{\mathscr{M}}
\safemath{\colN}{\mathscr{N}}
\safemath{\colO}{\mathscr{O}}
\safemath{\colP}{\mathscr{P}}
\safemath{\colQ}{\mathscr{Q}}
\safemath{\colR}{\mathscr{R}}
\safemath{\colS}{\mathscr{S}}
\safemath{\colT}{\mathscr{T}}
\safemath{\colU}{\mathscr{U}}
\safemath{\colV}{\mathscr{V}}
\safemath{\colW}{\mathscr{W}}
\safemath{\colX}{\mathscr{X}}
\safemath{\colY}{\mathscr{Y}}
\safemath{\colZ}{\mathscr{Z}}

\safemath{\opA}{\mathbb{A}}
\safemath{\opB}{\mathbb{B}}
\safemath{\opC}{\mathbb{C}}
\safemath{\opD}{\mathbb{D}}
\safemath{\opE}{\mathbb{E}}
\safemath{\opF}{\mathbb{F}}
\safemath{\opG}{\mathbb{G}}
\safemath{\opH}{\mathbb{H}}
\safemath{\opI}{\mathbb{I}}
\safemath{\opJ}{\mathbb{J}}
\safemath{\opK}{\mathbb{K}}
\safemath{\opL}{\mathbb{L}}
\safemath{\opM}{\mathbb{M}}
\safemath{\opN}{\mathbb{N}}
\safemath{\opO}{\mathbb{O}}
\safemath{\opP}{\mathbb{P}}
\safemath{\opQ}{\mathbb{Q}}
\safemath{\opR}{\mathbb{R}}
\safemath{\opS}{\mathbb{S}}
\safemath{\opT}{\mathbb{T}}
\safemath{\opU}{\mathbb{U}}
\safemath{\opV}{\mathbb{V}}
\safemath{\opW}{\mathbb{W}}
\safemath{\opX}{\mathbb{X}}
\safemath{\opY}{\mathbb{Y}}
\safemath{\opZ}{\mathbb{Z}}
\safemath{\opZero}{\mathbb{O}}
\safemath{\identityop}{\opI}

\safemath{\veca}{\bma}
\safemath{\vecb}{\bmb}
\safemath{\vecc}{\bmc}
\safemath{\vecd}{\bmd}
\safemath{\vece}{\bme}
\safemath{\vecf}{\bmf}
\safemath{\vecg}{\bmg}
\safemath{\vech}{\bmh}
\safemath{\veci}{\bmi}
\safemath{\vecj}{\bmj}
\safemath{\veck}{\bmk}
\safemath{\vecl}{\bml}
\safemath{\vecm}{\bmm}
\safemath{\vecn}{\bmn}
\safemath{\veco}{\bmo}
\safemath{\vecp}{\bmp}
\safemath{\vecq}{\bmq}
\safemath{\vecr}{\bmr}
\safemath{\vecs}{\bms}
\safemath{\vect}{\bmt}
\safemath{\vecu}{\bmu}
\safemath{\vecv}{\bmv}
\safemath{\vecw}{\bmw}
\safemath{\vecx}{\bmx}
\safemath{\vecy}{\bmy}
\safemath{\vecz}{\bmz}

\safemath{\veczero}{\bmzero}
\safemath{\vecone}{\bmone}
\safemath{\vecxi}{\bmxi}
\safemath{\veclambda}{\bmlambda}
\safemath{\vecmu}{\bmmu}
\safemath{\vectheta}{\bmtheta}
\safemath{\vecphi}{\bmphi}
\safemath{\vecdelta}{\bmdelta}

\safemath{\matA}{\bA}
\safemath{\matB}{\bB}
\safemath{\matC}{\bC}
\safemath{\matD}{\bD}
\safemath{\matE}{\bE}
\safemath{\matF}{\bF}
\safemath{\matG}{\bG}
\safemath{\matH}{\bH}
\safemath{\matI}{\bI}
\safemath{\matJ}{\bJ}
\safemath{\matK}{\bK}
\safemath{\matL}{\bL}
\safemath{\matM}{\bM}
\safemath{\matN}{\bN}
\safemath{\matO}{\bO}
\safemath{\matP}{\bP}
\safemath{\matQ}{\bQ}
\safemath{\matR}{\bR}
\safemath{\matS}{\bS}
\safemath{\matT}{\bT}
\safemath{\matU}{\bU}
\safemath{\matV}{\bV}
\safemath{\matW}{\bW}
\safemath{\matX}{\bX}
\safemath{\matY}{\bY}
\safemath{\matZ}{\bZ}
\safemath{\matzero}{\bmzero}

\safemath{\matDelta}{\bDelta}
\safemath{\matLambda}{\bLambda}
\safemath{\matPhi}{\bPhi}
\safemath{\matSigma}{\bSigma}
\safemath{\matOmega}{\bOmega}
\safemath{\matTheta}{\bTheta}

\safemath{\matidentity}{\matI}
\safemath{\matone}{\matO}

\safemath{\rnda}{A}
\safemath{\rndb}{B}
\safemath{\rndc}{C}
\safemath{\rndd}{D}
\safemath{\rnde}{E}
\safemath{\rndf}{F}
\safemath{\rndg}{G}
\safemath{\rndh}{H}
\safemath{\rndi}{I}
\safemath{\rndj}{J}
\safemath{\rndk}{K}
\safemath{\rndl}{L}
\safemath{\rndm}{M}
\safemath{\rndn}{N}
\safemath{\rndo}{O}
\safemath{\rndp}{P}
\safemath{\rndq}{Q}
\safemath{\rndr}{R}
\safemath{\rnds}{S}
\safemath{\rndt}{T}
\safemath{\rndu}{U}
\safemath{\rndv}{V}
\safemath{\rndw}{W}
\safemath{\rndx}{X}
\safemath{\rndy}{Y}
\safemath{\rndz}{Z}

\safemath{\rveca}{\bimA}
\safemath{\rvecb}{\bimB}
\safemath{\rvecc}{\bimC}
\safemath{\rvecd}{\bimD}
\safemath{\rvece}{\bimE}
\safemath{\rvecf}{\bimF}
\safemath{\rvecg}{\bimG}
\safemath{\rvech}{\bimH}
\safemath{\rveci}{\bimI}
\safemath{\rvecj}{\bimJ}
\safemath{\rveck}{\bimK}
\safemath{\rvecl}{\bimL}
\safemath{\rvecm}{\bimM}
\safemath{\rvecn}{\bimN}
\safemath{\rveco}{\bomO}
\safemath{\rvecp}{\bimP}
\safemath{\rvecq}{\bimQ}
\safemath{\rvecr}{\bimR}
\safemath{\rvecs}{\bimS}
\safemath{\rvect}{\bimT}
\safemath{\rvecu}{\bimU}
\safemath{\rvecv}{\bimV}
\safemath{\rvecw}{\bimW}
\safemath{\rvecx}{\bimX}
\safemath{\rvecy}{\bimY}
\safemath{\rvecz}{\bimZ}

\safemath{\rvecxi}{\bmxi}
\safemath{\rveclambda}{\bmlambda}
\safemath{\rvecmu}{\bmmu}
\safemath{\rvectheta}{\bmtheta}
\safemath{\rvecphi}{\bmphi}

\safemath{\rmatA}{\bimA}
\safemath{\rmatB}{\bimB}
\safemath{\rmatC}{\bimC}
\safemath{\rmatD}{\bimD}
\safemath{\rmatE}{\bimE}
\safemath{\rmatF}{\bimF}
\safemath{\rmatG}{\bimG}
\safemath{\rmatH}{\bimH}
\safemath{\rmatI}{\bimI}
\safemath{\rmatJ}{\bimJ}
\safemath{\rmatK}{\bimK}
\safemath{\rmatL}{\bimL}
\safemath{\rmatM}{\bimM}
\safemath{\rmatN}{\bimN}
\safemath{\rmatO}{\bimO}
\safemath{\rmatP}{\bimP}
\safemath{\rmatQ}{\bimQ}
\safemath{\rmatR}{\bimR}
\safemath{\rmatS}{\bimS}
\safemath{\rmatT}{\bimT}
\safemath{\rmatU}{\bimU}
\safemath{\rmatV}{\bimV}
\safemath{\rmatW}{\bimW}
\safemath{\rmatX}{\bimX}
\safemath{\rmatY}{\bimY}
\safemath{\rmatZ}{\bimZ}

\safemath{\rmatDelta}{\bimDelta}
\safemath{\rmatLambda}{\bimLambda}
\safemath{\rmatPhi}{\bimPhi}
\safemath{\rmatSigma}{\bimSigma}
\safemath{\rmatOmega}{\bimOmega}
\safemath{\rmatTheta}{\bimTheta}

\usepackage{amssymb}
\usepackage{amsfonts}
\usepackage{mathrsfs}
\usepackage{xspace}
\usepackage{bm}
\usepackage{fancyref}
\usepackage{textcomp}

\usepackage{multirow}
\usepackage{stmaryrd}

\newenvironment{textbmatrix}{	\setlength{\arraycolsep}{2.5pt}%
								\big[\begin{matrix}}{\end{matrix}\big]%
								\raisebox{0.08ex}{\vphantom{M}}}

\def\be{\begin{equation}}
\def\ee{\end{equation}}
\def\een{\nonumber \end{equation}}
\def\mat{\begin{bmatrix}}
\def\emat{\end{bmatrix}}
\def\btm{\begin{textbmatrix}}
\def\etm{\end{textbmatrix}}

\def\ba#1\ea{\begin{align}#1\end{align}}
\def\bas#1\eas{\begin{align*}#1\end{align*}}
\def\bs#1\es{\begin{split}#1\end{split}}
\def\bg#1\eg{\begin{gather}#1\end{gather}}
\def\bml#1\eml{\begin{multline}#1\end{multline}}
\def\bi#1\ei{\begin{itemize}#1\end{itemize}}

\newcommand{\lefto}{\mathopen{}\left}

\newcommand{\vecnorm}[1]{\lefto\lVert#1\right\rVert}		

\safemath{\dirac}{\delta}					
\safemath{\krond}{\dirac}					

\safemath{\upto}{\uparrow}
\safemath{\downto}{\downarrow}
\safemath{\iu}{j}							
\safemath{\ev}{\lambda}						
\safemath{\hilseqspace}{l^{2}}				
\newcommand{\banachfunspace}[1]{\setL^{#1}}	
\safemath{\hilfunspace}{\banachfunspace{2}}	

\safemath{\SNR}{\textit{SNR}} 				
\safemath{\PAR}{\textit{PAR}} 				
\safemath{\No}{N_0}							
\safemath{\Es}{E_s}							
\safemath{\Eb}{E_b}							
\safemath{\EbNo}{\frac{\Eb}{\No}}
\safemath{\EsNo}{\frac{\Es}{\No}}

\DeclareMathOperator{\CHop}{\ensuremath{\opH}} 
\safemath{\tvir}{\rndh_{\CHop}}				
\safemath{\tvtf}{\rndl_{\CHop}}				
\safemath{\spf}{\rnds_{\CHop}}				
\safemath{\bff}{H_{\CHop}}					

\safemath{\ircf}{r_{h}}						
\safemath{\tftvcf}{r_{s}}					
\safemath{\tfcf}{r_{l}}						
\safemath{\bfcf}{r_{H}}						

\safemath{\tcorr}{c_h}						
\safemath{\scf}{c_{s}}						
\safemath{\tfcorr}{c_{l}}					
\safemath{\fcorr}{c_{H}}						

\safemath{\mi}{I}							
\safemath{\capacity}{C}						

\safemath{\normal}{\mathcal{N}}			
\safemath{\jpg}{\mathcal{CN}}			
\safemath{\mchain}{\leftrightarrow}		

\safemath{\dB}{\,\mathrm{dB}}
\safemath{\dBm}{\,\mathrm{dBm}}
\safemath{\Hz}{\,\mathrm{Hz}}
\safemath{\kHz}{\,\mathrm{kHz}}
\safemath{\MHz}{\,\mathrm{MHz}}
\safemath{\GHz}{\,\mathrm{GHz}}
\safemath{\s}{\,\mathrm{s}}
\safemath{\ms}{\,\mathrm{ms}}
\safemath{\mus}{\,\mathrm{\text{\textmu}s}}
\safemath{\ns}{\,\mathrm{ns}}
\safemath{\ps}{\,\mathrm{ps}}
\safemath{\meter}{\,\mathrm{m}}
\safemath{\mm}{\,\mathrm{mm}}
\safemath{\cm}{\,\mathrm{cm}}
\safemath{\m}{\,\mathrm{m}}
\safemath{\W}{\,\mathrm{W}}
\safemath{\mW}{\, \mathrm{mW}}
\safemath{\J}{\,\mathrm{J}}
\safemath{\K}{\,\mathrm{K}}
\safemath{\bit}{\,\mathrm{bit}}
\safemath{\nat}{\,\mathrm{nat}}

\safemath{\define}{\triangleq}			

\safemath{\equivalent}{\sim}
\safemath{\distas}{\sim}					
\safemath{\sdiff}{\Delta}				

\safemath{\reals}{\mathbb{R}}
\safemath{\positivereals}{\reals_{+}}
\safemath{\integers}{\mathbb{Z}}
\safemath{\posint}{\integers_{+}}
\safemath{\naturals}{\mathbb{N}}
\safemath{\posnaturals}{\naturals_{+}}
\safemath{\complexset}{\mathbb{C}}
\safemath{\rationals}{\mathbb{Q}}

\newcommand*{\fancyrefapplabelprefix}{app}		
\newcommand*{\fancyrefthmlabelprefix}{thm}		
\newcommand*{\fancyreflemlabelprefix}{lem}		
\newcommand*{\fancyrefcorlabelprefix}{cor}		
\newcommand*{\fancyrefdeflabelprefix}{def}		
\newcommand*{\fancyrefproplabelprefix}{prop}		
\newcommand*{\fancyrefexmpllabelprefix}{exmpl}
\newcommand*{\fancyrefalglabelprefix}{alg}		
\newcommand*{\fancyreftbllabelprefix}{tbl}		

\frefformat{vario}{\fancyrefseclabelprefix}{Sec.~#1}
\frefformat{vario}{\fancyrefthmlabelprefix}{Thm.~#1}
\frefformat{vario}{\fancyreftbllabelprefix}{Tbl.~#1}
\frefformat{vario}{\fancyreflemlabelprefix}{Lem.~#1}
\frefformat{vario}{\fancyrefcorlabelprefix}{Cor.~#1}
\frefformat{vario}{\fancyrefdeflabelprefix}{Def.~#1}
\frefformat{vario}{\fancyreffiglabelprefix}{Fig.~#1}
\frefformat{vario}{\fancyrefapplabelprefix}{App.~#1}
\frefformat{vario}{\fancyrefeqlabelprefix}{(#1)}
\frefformat{vario}{\fancyrefproplabelprefix}{Prop.~#1}
\frefformat{vario}{\fancyrefexmpllabelprefix}{Ex.~#1}
\frefformat{vario}{\fancyrefalglabelprefix}{Alg.~#1}

\safemath{\dictab}{[\,\dicta\,\,\dictb\,]}

\safemath{\ysig}{\bmy}
\safemath{\ysighat}{\hat{\ysig}}
\safemath{\ysigdim}{M}
\safemath{\xsig}{\bmx}
\safemath{\xsigdim}{N}
\safemath{\nx}{n_x}
\safemath{\zsig}{\bmz}
\safemath{\zsigdim}{\ysigdim}
\safemath{\rsig}{\bmr}
\safemath{\Adict}{\bA}
\safemath{\Adicttilde}{\widetilde{\Adict}}
\safemath{\Adictdim}{\outputdim\times\xsigdim}
\safemath{\avec}{\bma}
\safemath{\avectilde}{\tilde{\avec}}
\safemath{\Bdict}{\bB}
\safemath{\Bdicttilde}{\widetilde{\Bdict}}
\safemath{\Cdict}{\bC}
\safemath{\cvec}{\bmc}
\safemath{\Ddict}{\bD}
\safemath{\Ddictdim}{\ysigdim\times\xsigdim}
\safemath{\dvec}{\bmd}
\safemath{\Ddicttilde}{\widetilde{\bD}}
\safemath{\Bonb}{\bB}
\safemath{\bvec}{\bmb}
\safemath{\Bonbdim}{\ysigdim\times\ysigdim}
\safemath{\noise}{\bmn}
\safemath{\noisedim}{\ysigim}
\safemath{\err}{\bme}
\safemath{\errdim}{\ysigdim}
\safemath{\errset}{\setE}
\safemath{\nerr}{n_e}
\safemath{\delop}{\bP_\errset}
\safemath{\delopc}{\bP_{{\errset}^c}}

\safemath{\cplxi}{\imath}
\safemath{\cplxj}{\jmath}

\safemath{\dict}{\matD}
\safemath{\inputdim}{N}		
\safemath{\outputdim}{M}		
\safemath{\sparsity}{S}	
\safemath{\inputdimA}{{N_a}}	
\safemath{\inputdimB}{{N_b}}	
\safemath{\elemA}{{n_a}}	
\safemath{\elemB}{{n_b}}	
\safemath{\resA}{\matR_a}	
\safemath{\resB}{\matR_b}	
\safemath{\subD}{\matS} 
\safemath{\subA}{\matS_a} 
\safemath{\subB}{\matS_b} 
\safemath{\dicta}{\matA} 	
\safemath{\dictb}{\matB} 	
\safemath{\hollowS}{H}
\safemath{\hollowA}{H_a}
\safemath{\hollowB}{H_b}
\safemath{\cross}{Z}
\safemath{\coh}{\mu_d}			
\safemath{\coha}{\mu_a}			
\safemath{\cohb}{\mu_b}			
\safemath{\mubs}{\nu}	
\safemath{\cohm}{\mu_m} 
\safemath{\dictset}{\setD}	
\safemath{\dictsetp}{\dictset(\coh,\coha,\cohb)}	
\safemath{\dictsetgen}{\dictset_\text{gen}}
\safemath{\dictsetgenp}{\dictsetgen(\coh)}
\safemath{\dictsetonb}{\dictset_\text{onb}}
\safemath{\dictsetonbp}{\dictsetonb(\coh)}

\safemath{\leftside}{U}
\safemath{\rightsideA}{R_a}
\safemath{\rightsideB}{R_b}

\safemath{\indexS}{\setI_S} 

\safemath{\na}{n_a}			
\safemath{\nb}{n_b}			
\safemath{\coeffa}{p_i}	
\safemath{\coeffb}{q_j}	
\safemath{\seta}{\setP}		
\safemath{\setb}{\setQ}     
\safemath{\setw}{\setW}	
\safemath{\setz}{\setZ}	
\safemath{\cola}{\veca}		
\safemath{\colb}{\vecb}		
\safemath{\cold}{\vecd}		
\safemath{\inputvec}{\vecx} 	
\safemath{\error}{\vece}	
\safemath{\noiseout}{\vecz} 	
\safemath{\inputvecel}{x}
\safemath{\inputveca}{\vecx_a}
\safemath{\inputvecb}{\vecx_b}
\safemath{\outputvec}{\vecy}	
\safemath{\lambdamin}{\lambda_{\mathrm{min}}}

\safemath{\elltwo}{\ell_2}
\safemath{\ellone}{\ell_1}
\safemath{\ellzero}{\ell_0}
\safemath{\ellinf}{\ell_\infty}
\safemath{\ellinftilde}{\ell_{\widetilde\infty}}
\safemath{\licard}{Z(\coh,\coha,\cohb)}
\safemath{\xsol}{\hat{x}}
\safemath{\xbord}{x_b}		
\safemath{\xstat}{x_s}		
\safemath{\xstatLone}{\tilde{x}_s}
\safemath{\order}{\mathcal{O}} 
\safemath{\scales}{\Theta} 
\safemath{\ones}{\mathbf{1}} 
\safemath{\zeroes}{\mathbf{0}} 
\safemath{\thlone}{\kappa(\coh,\cohb)} 
\safemath{\constoneA}{\delta} 
\safemath{\constoneB}{\epsilon} 
\safemath{\nlarge}{L}				   
\safemath{\sumlarge}{S_\nlarge}
\safemath{\maxlarger}{P_\nlarge}	   
\safemath{\Pzero}{\textrm{P0}}	
\safemath{\Pone}{\textrm{P1}}
\safemath{\vecfir}{\vecw}			 
\safemath{\vecsec}{\vecz}
\safemath{\elvecfir}{w}              
\safemath{\elvecsec}{z}				 
\safemath{\nlargefir}{n}
\safemath{\normout}{\gamma}
\safemath{\auxfun}{h}
\safemath{\supp}{\textrm{supp}}

\safemath{\indexa}{\ell}
\safemath{\indexb}{r}
\safemath{\indexc}{i}
\safemath{\indexd}{j}

\safemath{\project}{P}

\newcommand*{\fancyrefremarklabelprefix}{remark}
\frefformat{vario}{\fancyrefremarklabelprefix}{Remark~#1}

\newcommand{\PR}[1]{\ensuremath{\!\left[#1\right]}}
\newcommand{\PC}[1]{\ensuremath{\!\left(#1\right)}}
\newcommand{\chav}[1]{\ensuremath{\!\left\{#1\right\}}}
\newcommand{\PM}[1]{\ensuremath{\!\left|#1\right|}}

\usepackage{color}

\def\loss{\mathfrak{L}}

\IEEEoverridecommandlockouts 
\allowdisplaybreaks 


\title{CSI2Vec: Towards a Universal CSI Feature Representation for Positioning and Channel Charting\author{\IEEEauthorblockN{Victoria Palhares, Sueda Taner, and Christoph Studer}\thanks{V. Palhares, S. Taner, and C. Studer are with the Department of Information Technology and Electrical Engineering, ETH Zurich, Switzerland (palhares@iis.ee.ethz.ch, taners@iis.ee.ethz.ch, and studer@ethz.ch).}
\thanks{This work was supported in part by the Swiss National Science Foundation (SNSF) grant 200021\_207314 and by CHIST-ERA grant for the project CHASER (CHIST-ERA-22-WAI-01) through the SNSF grant 20CH21\_218704.}
\thanks{The authors thank Remcom \cite{remcom} for providing a license for the Wireless InSite ray-tracing software. The authors also thank Olav Tirkkonen and Maxime Guillaud for discussions on CSI2Vec.}}}

\begin{document}

\maketitle


\begin{abstract}
Natural language processing techniques, such as Word2Vec, have demonstrated exceptional capabilities in capturing semantic and syntactic relationships of text through vector embeddings. 
Inspired by this technique, we propose CSI2Vec, a self-supervised framework for generating universal and robust channel state information (CSI) representations tailored to CSI-based positioning (POS) and channel charting (CC). 
CSI2Vec learns compact vector embeddings across various wireless scenarios, capturing spatial relationships between user equipment positions without relying on CSI reconstruction or ground-truth position information.
We implement CSI2Vec as a neural network that is trained across various deployment setups (i.e., the spatial arrangement of radio equipment and scatterers) and radio setups (RSs) (i.e., the specific hardware used), ensuring robustness to aspects such as differences in the environment, the number of used antennas, or allocated set of subcarriers.
CSI2Vec abstracts the RS by generating compact vector embeddings that capture essential spatial information, avoiding the need for full CSI transmission or reconstruction while also reducing complexity and improving processing efficiency of downstream tasks. 
Simulations with ray-tracing and real-world CSI datasets demonstrate CSI2Vec’s effectiveness in maintaining excellent POS and CC performance while reducing computational demands and storage.

\end{abstract}


\section{Introduction} 
\label{sec:introduction}
Natural language processing (NLP) has demonstrated remarkable success in learning universal representations of text using vector embedding techniques such as Word2Vec\mbox{\cite{Mikolov2013a, Mikolov2013b}}. 
Inspired by such results, we propose leveraging similar representation learning techniques to learn universal vector embeddings for channel state information (CSI), thereby enabling a unified spatial representation of the wireless environment.
CSI inherently encodes rich spatial information about the physical surroundings, making it useful for various wireless sensing tasks, including mobility tracking, radio environment mapping, and object detection \cite{Rocamora2020,Wu2015,Wang2018,Vuckovic2021,Gonultas2022}.
Analogously to how words in NLP reflect semantic and syntactic relationships, CSI variations across different positions capture the underlying spatial structures of the radio environment. By mapping CSI to a latent space using vector embeddings, we can establish a universal CSI feature representation that captures relevant spatial dependencies and enables various sensing applications.

Recent studies have demonstrated the effectiveness of vector embeddings for wireless applications,  such as positioning (POS) \cite{Gonultas2022,Vieira2017,Garcia2017,Bast2020,Wang2017,Wu2013,Hsieh2019} and channel charting (CC) \cite{Studer2018, Taner2024,Stephan2024,Agostini2022,Deng2018}. 
These techniques exploit information captured by CSI to infer spatial relations, and by embedding CSI in a compact latent space, we can efficiently utilize a common representation to enable robust and accurate location-based sensing tasks. 
This unified framework enables diverse spatial inference tasks with improved generalization and reduced need for task-specific training.

\subsection{Contributions} 

\begin{table*}
\centering
\hspace{-1.0cm}
\resizebox{0.95\textwidth}{!}{
\begin{minipage}[c]{0.99 \textwidth}
\centering
\caption{Comparison of the proposed CSI2Vec approach with recent relevant prior art.}
\label{tbl:comparison_prior_art}
\begin{tabular}{@{}lcccccccc@{}}
\toprule
Properties & CSI2Vec & Huang\cite{Huang2019} & Alikhani\cite{Alikhani2024} & Ott\cite{Ott2024} & Foliadis\cite{Foliadis2023} & Salihu\cite{Salihu2024} & Jiang\cite{Jiang2025a} & Aboulfotouh\cite{Aboulfotouh2025} \\
\midrule
Training across deployment setups (DSs) & \green{$\checkmark$} & \red{$\times$} & \red{$\times$} & \red{$\times$} & \red{$\times$} & \red{$\times$} & \red{$\times$} & \red{$\times$} \\
Training across radio setups (RSs) & \green{$\checkmark$} & \red{$\times$} & \red{$\times$} & \red{$\times$} & \red{$\times$} & \red{$\times$} & \red{$\times$} & \red{$\times$}\\
Data augmentation (AUG) & \green{$\checkmark$} & \red{$\times$} & \red{$\times$} & \red{$\times$}  & \red{$\times$}  &  \green{$\checkmark$} & \red{$\times$} & \green{$\checkmark$}\\
Self-supervised training & \green{$\checkmark$} & \green{$\checkmark$} & \green{$\checkmark$} & \green{$\checkmark$} & \red{$\times$} & \green{$\checkmark$} & \green{$\checkmark$} & \green{$\checkmark$} \\
Suitable for OFDMA & \green{$\checkmark$} & \red{$\times$} & \green{$\checkmark$} & \red{$\times$} & \green{$\checkmark$} & \green{$\checkmark$} & \green{$\checkmark$} & \green{$\checkmark$} \\
Measured data & \green{$\checkmark$} & \red{$\times$} & \red{$\times$} & \green{$\checkmark$} & \red{$\times$} & \green{$\checkmark$} & \red{$\times$} & \green{$\checkmark$}\\
Accounts for hardware impairments & \green{$\checkmark$} & \green{$\checkmark$} & \red{$\times$} & \red{$\times$} & \red{$\times$} & \green{$\checkmark$} & \red{$\times$} & \red{$\times$} \\
\midrule
{Downstream tasks} & & & & & & \\
\midrule
CSI compression & \red{$\times$} & \green{$\checkmark$} & \green{$\checkmark$} & \green{$\checkmark$} & \red{$\times$} & \red{$\times$} & \red{$\times$} & \green{$\checkmark$} \\
Positioning (POS) & \green{$\checkmark$} & \green{$\checkmark$} & \red{$\times$} & \green{$\checkmark$} & \green{$\checkmark$} & \green{$\checkmark$} & \green{$\checkmark$} & \green{$\checkmark$} \\
Channel charting (CC) & \green{$\checkmark$} & \green{$\checkmark$} & \red{$\times$} & \red{$\times$} & \red{$\times$} & \red{$\times$} & \red{$\times$} & \red{$\times$}\\
\bottomrule
\end{tabular}
\end{minipage}}
\end{table*}

We propose CSI2Vec, a self-supervised framework for generating universal CSI feature representations tailored to POS and CC applications.
In this context, we draw an analogy between words and CSI samples,
as well as between various texts or books and different deployment setups (DSs) (i.e., the spatial arrangement of radio equipment and scatterers) and radio setups (RSs) (i.e., the specific hardware used).
Just as Word2Vec groups words with similar semantic and syntactic roles closely in a latent space, CSI2Vec maps CSI samples from nearby UE positions to vector embeddings near each other.
Moreover, similar to how words appear in different texts, CSI samples
are found across various DSs and RSs, further enhancing the model’s ability to capture spatial relationships.
To achieve this, we propose computing vector embeddings through self-
and semi-supervised learning (SEMI) across multiple distinctive scenarios. Each scenario involves unique DSs and RSs, incorporating variations in factors such as the number of receive antennas, subcarrier allocation, and environmental conditions. 
CSI2Vec can abstract the RSs by generating compact (i.e., low-dimensional) vector embeddings that preserve essential spatial information while avoiding the need for the entire CSI measurement.
Furthermore, the low-dimensional nature of these vector embeddings reduces computational complexity and storage requirements, which improves the processing efficiency of downstream tasks.
We validate the effectiveness of CSI2Vec by leveraging the learned universal CSI vector embeddings for the downstream tasks of CSI-based POS and CC, using channel vectors from both a commercial ray tracer \cite{remcom} and real-world measurements \cite{dataset-dichasus-cf0x}.

\subsection{Relevant Prior Art}
Existing approaches for constructing vector embeddings for CSI representation exhibit several limitations, particularly in terms of universality and adaptability across different scenarios. 
Some previous studies train models on specific datasets and individually for each scenario, resulting in vector embeddings that fail to generalize effectively across diverse DSs and RSs \cite{Alikhani2024,Jiang2025a,Foliadis2023,Ott2024,Salihu2024,Chen2022,Aboulfotouh2025,Guler2025}. 
Furthermore, some studies rely solely on simulated data, such as the DeepMIMO dataset \cite{Alkhateeb2019}, for training, without incorporating real-world measurements \cite{Alikhani2024,Foliadis2023,Jiang2025a,Guler2025}. This dependence on synthetic data limits the robustness of the learned representations towards hardware imperfections and highlights the need for approaches capable of learning across heterogeneous datasets to enable reliable POS and CC in diverse wireless scenarios. 
In contrast, CSI2Vec aims at learning universal vector embeddings for POS and CC through training across multiple heterogeneous datasets with diverse DSs and RSs, including not only simulated data but also real-world measurements from the DICHASUS dataset~\cite{dataset-dichasus-cf0x}.

A well-established approach in the literature involves generating vector embeddings with autoencoders (AEs) by minimizing the difference between the reconstructed and original CSI samples \cite{Alikhani2024,Ott2024,Aboulfotouh2025,Guler2025}. 
This approach is commonly used for CSI compression~\cite{Guo2022, Fan2024, Zeng2024, Wen2018, Wang2019, Lu2020, So2023, Liu2025, Zhang2023, Chatelier2024, Sattari2025, Jiang2025b}. 
However, as we demonstrate, representations for CSI compressions are not only unnecessary but can also be detrimental for CSI-based POS and CC applications, as the loss function of AEs does not explicitly enforce spatial relationships between CSI samples, which are crucial for location-based tasks.
Prior work \cite{Huang2019} has shown that pairwise vector embedding constraints are imperative in AEs to construct meaningful POS maps and channel charts, as traditional architectures fail to preserve geometric properties in CSI vector embeddings.
In contrast, CSI2Vec does not rely on CSI reconstruction, but on establishing spatial relationships between CSI samples from diverse scenarios, making it well-suited for learning vector embeddings for POS and CC.

Alternative approaches for generating vector embeddings focus on establishing relationships within variations of the channel at the same UE position. 
For example, in \cite{Jiang2025a}, the objective is to minimize the distance between the vector embeddings of the CSI and of the complex impulse response (CIR) obtained from the same channel realization while maximizing the distance between vector embeddings from different channel realizations.
Similarly, some methods enforce similarity constraints between noisy versions of the same channel realization \cite{Salihu2024,Guler2025} or between temporal variations of the channel for the same access point (AP)-UE pair \cite{Liu2024a,Ismayilov2024}. 
Furthermore, \cite{Chaaya2024} proposes predicting future vector embeddings from a given {CSI} sample using UE velocity as side information.
In contrast, our approach establishes relationships between CSI samples across multiple scenarios rather than focusing on variations of the same CSI sample, improving generality and making it more suitable for spatially dependent applications.

Finally, some methods rely on ground-truth position information \cite{Foliadis2023,Foliadis2024a,Foliadis2024b} or camera data \cite{Chen2022} to generate meaningful vector embeddings.
However, such side information is often unavailable in practical scenarios, limiting the applicability of these techniques. 
In contrast, we learn vector embeddings through self-supervised learning solely relying on CSI features.

In \fref{tbl:comparison_prior_art}, we summarize the key differences between our proposed approach and the most relevant prior work. 

\subsection{Notation}
Column vectors are denoted by lowercase boldface letters; matrices and tensors are denoted by uppercase boldface letters; sets are denoted by uppercase calligraphic letters.
The transpose is denoted by the superscript $(\cdot)^T$.
The column-wise vectorization of $\bA$ is denoted by $\mathrm{vec}(\bA)$. 
For a vector~$\bma$, the Euclidean norm is $\|\bma\|$ and the entry-wise absolute value is $|\bma|$. 
The cardinality of a set~$\setA$ is $|\setA|$. 
The operators $\Re \{\mathrm{a}\}$ and $\Im \{\mathrm{a}\}$ extract the real and imaginary part of $a \in \opC$, respectively.
The operator $(x)^+=\max\{x,0\}$ is the rectified linear unit (ReLU).

\subsection{Paper Outline}
The rest of the paper is organized as follows. 
\fref{sec:csi2vec} introduces CSI2Vec and describes the pipeline used to obtain universal CSI {feature} representations. 
\fref{sec:downstream_tasks} presents two example downstream tasks that showcase the potential of CSI2Vec. 
\fref{sec:methods} details the proposed methods, baselines, and performance metrics. 
\fref{sec:results} showcases results for multiple {scenarios with diverse DSs and RSs}. 
\fref{sec:conclusions} concludes. 


\section{CSI2Vec: Universal CSI Representation for Positioning (POS) and Channel Charting (CC)} \label{sec:csi2vec}
\begin{figure*}[ht]
\centering
\includegraphics[width=\textwidth]{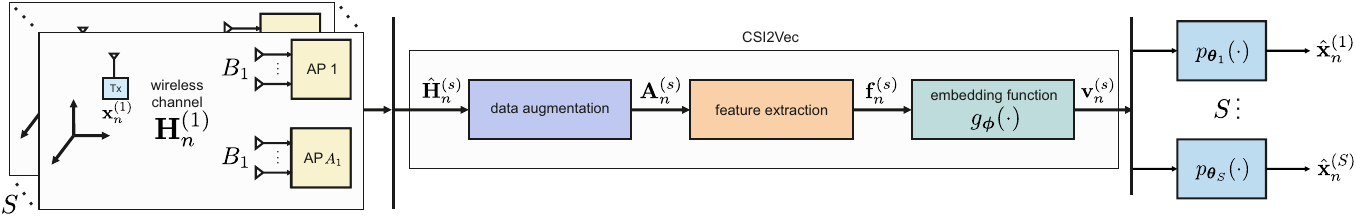}
\caption{CSI2Vec pipeline overview: 
A UE transmits pilot signals from $N_s$ ground-truth positions $\bmx_n^{(s)}$ to $A_s$ APs over the wireless channel $\mathbf{H}_n^{(s)}$. 
The CSI2Vec framework first processes the estimated CSI tensor $\hat\bH_n^{(s)}$ by standardizing its dimensions and applying data augmentation (AUG).
Next, the framework transforms the CSI from the subcarrier domain to the delay domain, followed by the truncation of channel taps in the delay domain. 
The processed CSI is then used to extract features, which are later fed into the embedding function $g_{\boldsymbol{\phi}}$, which is implemented by a multi-layer perceptron (MLP).  
The resulting vector embeddings enable scenario-specific (SCS) POS and CC  downstream tasks, accomplished by the functions~$p_{\boldsymbol{\theta}_s}$.}
\label{fig:pipeline}
\end{figure*}

We now describe the system model capturing different scenarios with diverse DSs and RSs, and we detail the processing pipeline of CSI2Vec as illustrated in \fref{fig:pipeline}. 

\subsection{System Model} \label{sec:system_model}
We consider $S$ different scenarios of uplink multiple-input multiple-output (MIMO) communication systems, i.e.,~$S$ distinct DSs and RSs. 
In the $s$th scenario, one or multiple UEs equipped with $U_s$ antennas each transmit pilots to $A_s$ distributed {APs} with $B_s$ antennas each. 
We consider frequency-selective channels divided into $W_s$ frequency-flat subcarriers using orthogonal frequency division multiplexing (OFDM).

For a given scenario $s$, $s\in\setS=\{1,\dots,S\}$, we assume that one UE transmits pilots from $N_s$ ground-truth positions $\bmx^{(s)}_{n}\in\opR^2$ at timestamps $t^{(s)}_{n}$ for $n\in\setN_s=\{1,\dots,N_s\}$, where $\setN_s$ is the set of all UE positions in scenario $s$. 
For each $n$th pilot transmission, the AP $a_s\in\setA_s=\{1,\dots,A_s\}$ estimates the associated CSI vector $\hat\bmh^{(s)}_{n,a_s,u_s,w_s}\in\opC^{B_s}$ on a subcarrier $w_s\in\setW_s=\{1,\dots,W_s\}$. 
We construct the CSI tensor {$\hat\bH^{(s)}_{n}\in\opC^{A_s \times B_s \times U_s \times W_s}$} by combining the vectors $\hat\bmh^{(s)}_{n,a_s,u_s,w_s}$ together for each one of the $n$th pilot transmissions in the $s$th scenario.

\subsection{Data Augmentation (AUG)} \label{sec:data_augmentation}
In CSI2Vec, our goal is to create universal CSI representations, robust to different scenarios with diverse DSs and RSs. 
Since we aim to train our vector embeddings across scenarios, it is crucial that our input features have the same dimensions. 
For this reason, we take the tensor {$\hat\bH^{(s)}_{n}\in\opC^{A_s \times B_s \times U_s \times W_s}$} from scenario~$s$ and zero-pad the tensor so that its new dimensions are {$\underline{\hat\bH}^{(s)}_{n}\in\opC^{A_{\text{max}} \times B_{\text{max}} \times U_{\text{max}} \times W_{\text{max}}}$}, where 
\begin{align} \label{eq:max_values}
    A_{\text{max}} &\triangleq \underset{s \in \{1,\ldots,S\}}{\max} A_s, \ B_{\text{max}} \triangleq \underset{s \in \{1,\ldots,S\}}{\max} B_s, \notag \\ U_{\text{max}} &\triangleq \underset{s \in \{1,\ldots,S\}}{\max} U_s, \text{ and } W_{\text{max}} \triangleq \underset{s \in \{1,\ldots,S\}}{\max} W_s.
\end{align}

Once all of our CSI tensors have the same dimensions across scenarios, we apply data augmentation (AUG) by randomly removing, i.e., zeroing out, certain entries and adding noise. 
This approach enables us to generate additional variations of our ray-traced and measured scenarios by imitating the removal of transmit and receive antennas as well as different subcarrier allocations to mimic orthogonal frequency-division multiple access (OFDMA), and accounting for channel estimation errors. 

\subsubsection{Removing Transmit and Receive Antennas}
To ensure robustness to different RSs, i.e., hardware with a different number of transmit and receive antennas, we randomly deactivate up to $U_s-1$ and $B_s-1$ antennas at the UE and AP sides, preventing the complete removal of the UE or of any AP.

\subsubsection{Removing Subcarrier Bands} \label{sec:removal_subcarrier_bands}

To ensure compatibility with OFDMA-based systems, we randomly remove subcarrier bands from each $n$th CSI sample from each scenario $s$, where $n \in \setN_s$. 
We implement this by fixing a percentage $q$ of subcarriers to be removed from all scenarios. 
Then, for each CSI sample, we randomly select a subcarrier index and remove its neighboring subcarriers based on the percentage $q$.

\subsubsection{Adding Noise} \label{sec:adding_noise}
To ensure robustness to channel estimation errors, we add noise to the non-zero entries of the CSI tensor after removing transmit/receive antennas and subcarrier bands. 
We set the noise variance $N_0$ such that the highest signal-to-noise ratio (SNR) per AP remains within a specified range.

After these data augmentation steps, the augmented CSI tensor is given by $\bA^{(s)}_{n}\in\opC^{A_{\text{max}} \times B_{\text{max}} \times U_{\text{max}} \times W_{\text{max}}}$.
This step enables us to generate rich and diverse datasets that account for different RSs on both the AP and UE sides.

\subsection{Feature Extraction}
\label{sec:feature_extraction}

Our goal is to arrive at compact feature representations suitable for diverse scenarios with different DSs and RSs. 
To this end, we extract CSI features, starting by extracting the delay-domain representation using an inverse discrete Fourier transform (IDFT) over the $W_{\text{max}}$ subcarriers in our CSI tensor, $\bA^{(s)}_{n}\in\opC^{A_{\text{max}} \times B_{\text{max}} \times U_{\text{max}} \times W_{\text{max}}}$. 
This process yields $\hat{\bA}^{(s)}_{n}\in\opC^{A_{\text{max}} \times B_{\text{max}} \times U_{\text{max}} \times C_{\text{max}}}$, where {$C_{\text{max}} = W_{\text{max}}$} is the maximum number of channel taps. 
As we are primarily interested in the first few taps containing the highest receive power, we truncate the maximum number of channel taps to {$C^{'}_{\text{max}} < C_{\text{max}}$}. 
This results in the CSI tensor {$\tilde{\bA}^{(s)}_{n}\in\opC^{A_{\text{max}} \times B_{\text{max}} \times U_{\text{max}} \times C^{'}_{\text{max}}}$}, which has lower dimension than $\hat{\bA}^{(s)}_{n}$.

Since CSI2Vec is designed for POS and CC tasks, we are interested in removing small-scale fading features while maintaining large-scale fading properties~\cite{Vieira2017,Studer2018}. 
To this end, we adopt the CSI feature extraction approach from~\cite{lei19siamese} that extracts large-scale fading properties while being robust to hardware impairments (e.g., phase offsets between APs)~\cite{ferrand2021} and UE transmit power variations.
We compute CSI features~as 
\begin{align}
    \bmf_n^{(s)} = \frac{\PM{\mathrm{vec}\PC{\tilde{\bA}^{(s)}_{n}}}}{\vecnorm{\mathrm{vec}\PC{\tilde{\bA}^{(s)}_{n}}}} \in \opR^{D},
\end{align}
where $D = A_{\text{max}} B_{\text{max}} U_{\text{max}} C^{'}_{\text{max}}$. 
The set of vectors~$\{\bmf^{(s)}_{n}\}_{n\in \setN_{s}, s \in \setS}$ forms the CSI feature dataset. 

\subsection{Learning Universal Vector Embeddings from CSI features}
From this CSI feature dataset, we now learn the CSI2Vec function $g_{\boldsymbol{\phi}}: \opR^D\to\opR^{D'}$ that maps CSI features $\bmf_n^{(s)} \in \opR^{D}$ across scenarios to compact vector embeddings {$\bmv_n^{(s)} \in \opR^{D'}$}:
\begin{align} \label{eq:vec_embed}
    \bmv^{(s)}_{n} = g_{\boldsymbol{\phi}}\PC{\bmf^{(s)}_{n}}.
\end{align}
We implement $g_{\boldsymbol{\phi}}$ as a neural network with weights and biases described by $\boldsymbol{\phi}$. 
Various methods for learning vector embeddings from wireless channels rely on CSI compression using a mean squared error (MSE) loss, implemented with transformers \cite{Ott2024,Alikhani2024,Salihu2024,Leng2025,Guler2025} or convolutional neural networks (CNNs) \cite{Foliadis2023,Chen2022}.
In contrast, since CSI reconstruction is unnecessary for POS and CC, we can utilize a simple multi-layer perceptron (MLP) with a contrastive (triplet) loss~\cite{Schroff2015} to generate compact vector embeddings from CSI features from multiple scenarios with diverse DSs and RSs. 

\subsection{Self-Supervised CSI2Vec}
\label{sec:self_sup_csi2vec}
Since we aim to find universal CSI feature representations, we train $g_{\boldsymbol{\phi}}$ across multiple scenarios. 
To mimic the reasoning of Word2Vec, where words that have similar semantic and syntactic functions have similar vector embeddings and words that have distinct semantic and syntactic functions have dissimilar vector embeddings, we employ a triplet loss~\cite{ferrand2021} so that CSI features with close timestamps have similar vector embeddings and CSI features with far timestamps have dissimilar vector embeddings. 
Different from what has been done for triplet-based CC~\cite{ferrand2021,Taner2024}, we also form triplets across scenarios.
Therefore, to determine whether a sample is close or far in time to the anchor sample, we do the following: 
(i) if a sample is close in time to the anchor sample in the same scenario $s$, then they will have similar vector embeddings, 
(ii) if a sample is far in time to the anchor sample in the same scenario $s$, then they will have dissimilar vector embeddings, and (iii) if a sample comes from a different scenario $s'$ than the {scenario $s$} of the anchor sample, then they will have dissimilar vector embeddings. 
In summary, for a given anchor sample, the nearby sample can only be picked within the same scenario $s$ if it is close in time.
Then, for the same anchor sample, the far sample can be picked either within the same scenario $s$ if it is far in time or from another scenario $s' \neq s$.

The equations that describe this strategy are as follows. We define the set of all vector embeddings from all scenarios as 
\begin{align}
    \setV=\bigcup_{s=1}^S \chav{\bmv_n^{(s)}}_{n=1}^{N_s}.
\end{align}
We define the set of triplets coming from the same scenario as
\begin{align} \label{eq:triplets_same_scenario}
\setT_{\text{IN}} &= 
\bigcup_{s\in\setS}\Big\{\PC{\bmv_n^{(s)},
\bmv_c^{(s)},\bmv_f^{(s)}} \in \setV^3 :0 < \PM{t^{(s)}_n-t_c^{(s)}} \notag\\ 
&\qquad \qquad \leq T_{\mathrm{c}} <\PM{t^{(s)}_n-t_f^{(s)}}\Big\},
\end{align}
and the set of triplets across scenarios as
\begin{align} \label{eq:triplets_diff_scenario}
\setT_{\text{OUT}} &= 
\bigcup_{(s,s')\in\setS^2, s\neq s'} \Big\{\PC{\bmv_n^{(s)},
\bmv_c^{(s)},\bmv_f^{(s')}} \in \setV^3: \notag \\
&\qquad \qquad \qquad \qquad 0 < \PM{t^{(s)}_n-t_c^{(s)}} \leq T_{\mathrm{c}}\Big\}.
\end{align}
We define the joint set of triplets as
\begin{align}
    \setT = \setT_{\text{IN}} \cup \setT_{\text{OUT}}.
\end{align}
In \fref{eq:triplets_same_scenario} and \fref{eq:triplets_diff_scenario}, $T_{\mathrm{c}}>0$ is a parameter that determines whether two samples are close or far in time. 
The anchor sample has a vector embedding $\bmv_n^{(s)}$ and timestamp $t^{(s)}_n$. 
Similarly, the close sample has $\bmv_c^{(s)}$ and $t^{(s)}_c$, while the far sample has $\bmv_f^{(s)}$ and $t^{(s)}_f$ (or $\bmv_f^{(s')}$ and $t^{(s')}_f$ if the sample belongs to a different scenario).
Given that $t^{(s)}_n$ is closer to $t^{(s)}_c$ than $t^{(s)}_f$, we can also assume that the Euclidean distance between the vector embeddings $\bmv_n^{(s)}$ and $\bmv_c^{(s)}$ is smaller than that of $\bmv_n^{(s)}$ and $\bmv_f^{(s)}$. 
Moreover, we assume that the Euclidean distance between two vector embeddings from the same scenario should be smaller than that of two different scenarios.
We describe the resulting triplet loss as follows: 
\begin{align} \label{eq:loss_t}
\loss_{\mathrm{t}} = \frac{1}{|\setT|} \! \sum_{(\bmv_{n},\bmv_{c},\bmv_{f})\in\setT} \!\! \big(& \vecnorm{\bmv_{n}- \bmv_{c}} \notag\\
& - \vecnorm{\bmv_{n}- \bmv_{f}} + M_{\mathrm{t}}\big)^+.
\end{align}
Here, the margin parameter~$M_{\mathrm{t}}\geq0$ ensures that $\bmv_n$ is at least~$M_{\mathrm{t}}$ closer to $\bmv_c$ than to $\bmv_f$.

\subsection{Semi-Supervised CSI2Vec (SEMI)}
\label{sec:semi_sup_csi2vec}

To improve the spatial relationships contained in the vector embeddings, we can exploit the fact that we may have ground-truth  {UE position} information from some scenarios but not from all. 
Assume that we have ground-truth UE position information from the scenarios in $\setS' \subset \setS$. 
For SEMI, we  learn a scenario-specific (SCS) POS function $\tilde{p}_{\boldsymbol{\delta}_{s'}}: \opR^{D'}\to\opR^{d}$, for $s' \in \setS'$, that maps vector embeddings $\bmv_n^{(s')}$ to estimated UE positions $\hat{\bmx}_n^{(s')} \in \opR^d$ by 
\begin{align} \label{eq:estimate_pos_sup}
    \hat{\bmx}_n^{(s')} = \tilde{p}_{\boldsymbol{\delta}_{s'}}\PC{\bmv_n^{(s')}}.
\end{align}
We implement $\tilde{p}_{\boldsymbol{\delta}_{s'}}$ as an MLP with weights and biases described by $\boldsymbol{\delta}_{s'}$, specifically for a scenario $s'$. 
We learn the SCS POS function $\tilde{p}_{\boldsymbol{\delta}_{s'}}$ with an MSE loss given by
\begin{align}\label{eq:loss_sup}
    \loss_{\text{SUP}} = \sum_{s' \in \setS'} \frac{1}{|\setN_{s'}|} \sum_{n\in\setN_{s'}}  \vecnorm{\hat{\bmx}_n^{(s')} - \bmx_n^{(s')}}^2,
\end{align}
where $\hat{\bmx}_n^{(s')}$ is given by \fref{eq:estimate_pos_sup}. 
Then, our semi-supervised loss is $\loss_{\text{SEMI}} = \loss_{\text{SUP}} +\loss_{t}$,\footnote{Each loss can be assigned a different weight; however, for the sake of simplicity, we weight them equally in this work.\label{footnote_loss_weights}}
where $\loss_{t}$ is given by \fref{eq:loss_t}.
Our goal with this method is to use all of the information that might be available on some scenarios but not in others in order to learn  vector embeddings with improved POS and CC performance.


\section{Downstream Tasks} \label{sec:downstream_tasks}
We now present two downstream tasks for CSI2Vec: SCS POS and SCS CC.
For these downstream tasks, we can learn a scenario-specific function $p_{\boldsymbol{\theta}_s}: \opR^{D'}\to\opR^{d}$ (for scenario $s$) that maps vector embeddings $\bmv_n^{(s)} \in \opR^{D'}$ to estimated {UE} positions for POS and pseudo-positions for CC as
\begin{align} \label{eq:pseudo_ue_pos}
    \hat{\bmx}_n^{(s)} = p_{\boldsymbol{\theta}_s}\PC{\bmv_n^{(s)}} \in \opR^d. 
\end{align}
All parameters describing the function $p_{\boldsymbol{\theta}_s}$ are contained in~$\boldsymbol{\theta}_s$.

\subsection{Task I: Scenario-Specific (SCS) Positioning (POS)}
\label{sec:positioning}

We implement $p_{\boldsymbol{\theta}_s}$ in \fref{eq:pseudo_ue_pos} as an MLP with weights and biases described by $\boldsymbol{\theta}_s$, {for each} scenario $s$. 
We learn the SCS POS function  $p_{\boldsymbol{\theta}_s}$ with an MSE loss as
\begin{align}\label{eq:loss_mse}
    \loss_{\text{POS}_s} = \frac{1}{|\setN_s|} \sum_{n\in\setN_s}  \vecnorm{\hat{\bmx}_n^{(s)} - \bmx_n^{(s)}}^2,
\end{align}
where $\hat{\bmx}_n^{(s)}$ is given by \fref{eq:pseudo_ue_pos}. 
This loss function relies on ground-truth UE positions for all training samples, which may be impractical in real-world scenarios. 
Here, POS is used solely as an example to demonstrate the capabilities of CSI2Vec.

\subsection{Task II: Scenario-Specific (SCS) Channel Charting (CC)} \label{sec:cc}
There are multiple methods available for CC. In this work, we consider two approaches: Siamese network (CC-SN) \cite{lei19siamese} and principal component analysis (CC-PCA) \cite{Pearson1901,Hotteling1933}.

\subsubsection{Siamese Network (CC-SN)} \label{sec:sammons_mapping}
We implement $p_{\boldsymbol{\theta}_s}$ in \fref{eq:pseudo_ue_pos} as a neural network with weights and biases described by $\boldsymbol{\theta}_s$, for each scenario $s$. 
We learn the SCS CC function $p_{\boldsymbol{\theta}_s}$ with a ``Siamese'' loss \cite{Bromley1993} given by
\begin{align}\label{eq:loss_siamese}
    \loss_{\text{SM}_s} \! = \frac{1}{|\setN_s|} \sum_i^{N_s} 
    \sum_{j=i+1}^{N_s} \! \gamma_{i,j} \PC{\vecnorm{\hat{\bmx}^{(s)}_{i} - \hat{\bmx}^{(s)}_{j}}\!-\!\vecnorm{\bmv^{(s)}_{i}-\bmv^{(s)}_{j}}}^2\!\!,
\end{align}
where $\hat{\bmx}^{(s)}_{i}$ and $\hat{\bmx}^{(s)}_{j}$ are given in \fref{eq:pseudo_ue_pos}, $\bmv^{(s)}_{i}$ and $\bmv^{(s)}_{j}$ are given in \fref{eq:vec_embed}, and $\gamma_{i,j} = \vecnorm{\hat{\bmx}^{(s)}_{i} - \hat{\bmx}^{(s)}_{j}}^{-1}$, $i,j \in \setN_s$. 
Note that this method is a parametric extension of Sammon's mapping \cite{Sammon1969}.

\subsubsection{Principal Component Analysis (CC-PCA)} \label{sec:pca}
To obtain a channel chart through principal component analysis (PCA), we do the following:
We collect all vector embeddings $\bmv_n^{(s)}$ and concatenate them in a matrix $\bV_{s} = \PR{\bmv_1^{(s)},\dots,\bmv_{N_s}^{(s)}} \in \opR^{D'\times N_s}$. 
Then, we get $\bar{\bV}_{s}$ by normalizing each row of $\bV_{s}$ to have zero empirical mean. 
Next, we compute the eigenvalue decomposition of the empirical covariance matrix of $\bar{\bV}_{s}$ such that $\bar{\bV}_{s}\bar{\bV}^T_{s} = \bU \boldsymbol{\Sigma} \bU^T$, where $\bU, \boldsymbol{\Sigma} \in \opR^{D' \times D'}$. 
The diagonal entries of $\boldsymbol{\Sigma}$ are sorted in descending order.
To reduce the dimensions of our vector embeddings to $d$, we pick the eigenvectors corresponding to the $d$ largest eigenvalues, i.e., $\bmu_1,\dots,\bmu_d$ and multiply $\bU_d = \PR{\bmu_1,\dots,\bmu_d} \in \opR^{D' \times d}$ by the {vector $\bmv_n^{(s)}$} resulting in the $d$-dimensional vector {$\hat{\bmx}_n^{(s)} = \bU_d^T \bmv_n^{(s)}$}.

Unlike {POS (cf. \fref{sec:positioning})}, both methods, CC-SN and CC-PCA, are completely self-supervised and do not rely on any ground-truth UE positions.
Here, CC is used as another example to demonstrate the effectiveness of CSI2Vec.


\section{Proposed Methods, Baselines, and Performance~Metrics} \label{sec:methods}
We now present our proposed methods and a set of baselines, as well as the performance metrics used in our evaluation.  

\subsection{Proposed Methods}
We highlight that all of our proposed methods are trained across vastly different scenarios, including channel vectors obtained from ray-tracing \cite{remcom} and measurements \cite{dataset-dichasus-cf0x}, as well as from outdoor and indoor scenarios. 

We implement all of the proposed CSI2Vec functions, $g_{\boldsymbol{\phi}}$, as an MLP with $\{D,32,D'\}$ activations per layer, i.e., $D$ is the dimension of the input layer, $32$ is the number of activations in the hidden layer, and $D'$ is the dimension of the output layer. From now on, we use this notation to describe the number of activations per layer in an MLP.
We train the CSI2Vec MLP across multiple scenarios, and we pick the hyperparameter $M_t$ from \fref{eq:loss_t} for all scenarios jointly.
In this work, we set $M_t = 10$. Even though we train the triplet loss considering multiple scenarios simultaneously, the definition of whether a sample is close or far in time within a scenario is SCS. 
Therefore, in \fref{sec:results}, we define $T_c$ from \fref{eq:triplets_same_scenario} and \fref{eq:triplets_diff_scenario} for each scenario separately. 

We implement the SCS POS function $p_{\boldsymbol{\theta}_s}$ (cf. \fref{sec:positioning}) and the SCS CC function $p_{\boldsymbol{\theta}_s}$ (cf. \fref{sec:cc}), for every scenario $s$ as an MLP with $\{D',12,8,6,4,d\}$ activations per layer, except in the case of CC-PCA, which we detail in \fref{sec:pca}.

For all MLPs, all layers except for the last one use {ReLU} activations. We initialize all neural network weights using Glorot~\cite{Glorot2010} and train the network with the Adam optimizer.

\subsubsection{CSI2Vec Without Augmentation (CSI2Vec)}

As our first method, we present CSI2Vec without AUG. This method is described in the pipeline of \fref{sec:csi2vec}, by skipping the AUG stage described in \fref{sec:data_augmentation}. 
Therefore, in this case, we assume that $\bA_n^{(s)} = \underline{\hat\bH}_n^{(s)}$. 
This method considers only the self-supervised CSI2Vec loss function given in \fref{sec:self_sup_csi2vec}. 

\subsubsection{CSI2Vec With Augmentation (CSI2Vec-AUG)} \label{sec:method_csi2vec}
As our second method, we present CSI2Vec-AUG. 
This method is described in the pipeline of \fref{sec:csi2vec}, now considering AUG detailed in \fref{sec:data_augmentation}. 
This method considers only the self-supervised CSI2Vec loss function given in \fref{sec:self_sup_csi2vec}.

\subsubsection{CSI2Vec With Semi-Supervised Learning (CSI2Vec-AUG-SEMI)}
\label{sec:csi2vec_SEMI}
As our third method, we present CSI2Vec-AUG-SEMI. 
This method is detailed in the pipeline described in \fref{sec:csi2vec}, now considering AUG detailed in \fref{sec:data_augmentation} and both self-supervised and semi-supervised CSI2Vec loss functions from \fref{sec:self_sup_csi2vec} and \fref{sec:semi_sup_csi2vec}, respectively. 

\subsection{Baseline Methods}
For a fair comparison with existing techniques and to demonstrate the efficacy of CSI2Vec, we consider the following baseline methods. 

\subsubsection{Scenario-Specific End-to-End Training (SCS-EE)}
As the first baseline method, we use fully-supervised SCS  {end-to-end  (EE) training}. 
The goal of this baseline is to show the best possible POS and CC performance when training \emph{separately} for each scenario, with specific DSs and RSs.

For this method, we assume that the features are computed considering the steps described in \fref{sec:data_augmentation} to \fref{sec:feature_extraction}, solely skipping the AUG described in \fref{sec:data_augmentation}. 
Therefore, in this case, we assume that {$\bA_n^{(s)} = \underline{\hat\bH}_n^{(s)}$}.

After having computed the CSI features, we follow the procedure of the method described in \fref{sec:downstream_tasks}, where we can learn either an SCS POS or a CC function $p_{\boldsymbol{\Theta}_s}: 
\opR^{D}\to\opR^{d}$ for each scenario $s$, that maps features 
$\bmf_n^{(s)} \in \opR^{D}$ to pseudo-UE positions $\hat{\bmx}_n^{(s)} \in \opR^{d}$, as given by
\begin{align} \label{eq:estimate_pos_ee}
    \hat{\bmx}_n^{(s)} = p_{\boldsymbol{\Theta}_s}\PC{\bmf_n^{(s)}}.
\end{align}
We implement the SCS POS function, $p_{\boldsymbol{\Theta}_s}$ (cf. \fref{sec:positioning}) and the SCS CC function $p_{\boldsymbol{\Theta}_s}$ (cf. \fref{sec:cc}), for every scenario $s$ as an MLP with $\{D,320,160,80,40,20,10,5,d\}$ activations per layer, except in the case of CC with PCA, which we detail in \fref{sec:pca}.
For all of the MLPs, all layers except for the last one have ReLU activations. We initialize all weights using Glorot initialization \cite{Glorot2010} and train the MLP with Adam.

\subsubsection{Autoencoder Without Augmentation (AE)} \label{sec:ae_without_aug}
As the second baseline method, we use an AE \cite{Rumelhart1986,Hinton2006,VanderMaaten2009,Baldi2012}, as used for CSI compression \cite{Guo2022, Fan2024, Zeng2024, Wen2018, Wang2019, Lu2020, So2023, Liu2025, Zhang2023, Chatelier2024,Sattari2025} and CC \cite{Huang2019,Studer2018}, without AUG, that aims at minimizing the difference between the reconstructed and the original CSI. 
The goal of this baseline is to evaluate the POS and CC performance of AE architectures commonly used for learning vector embeddings \cite{Alikhani2024,Ott2024,Aboulfotouh2025,Guler2025}, but not specifically tailored to POS and CC applications.

For this method, it is crucial to note that we must preserve small-scale fading aspects to enable CSI \emph{reconstruction}, unlike the CSI feature transformations discussed in \fref{sec:feature_extraction}. Therefore, we only apply the zero-padding technique introduced in  \fref{sec:data_augmentation} and the delay-domain truncation described in \fref{sec:feature_extraction}, i.e., $\bA_n^{(s)} = \underline{\hat\bH}_n^{(s)}$.
We compute the raw CSI vector $\bmf_n^{(s)} \in \opR^{D''}$ as follows:
\begin{align} \label{eq:feature_autoencoder}
    \bmf_n^{(s)} = \textstyle
    \begin{bmatrix}
        \Re \chav{\mathrm{vec}\PC{\tilde{\bA}_n^{(s)}}} \\
        \Im \chav{\mathrm{vec}\PC{\tilde{\bA}_n^{(s)}}} \\
    \end{bmatrix},
\end{align} 
where $D'' = 2A_{\text{max}} B_{\text{max}} U_{\text{max}} C^{'}_{\text{max}}$.
After having computed the raw CSI vector $\bmf_n^{(s)}$, we learn an encoding function $g_{\boldsymbol{\varphi}}: \opR^{D''}\to\opR^{D'}$ and a decoding function $g_{\boldsymbol{\Phi}}: \opR^{D'}\to\opR^{D''}$, that maps {the raw CSI vector} $\bmf_n^{(s)} \in \opR^{D''}$ across scenarios to reconstructed raw CSI vector $\hat{\bmf}_n^{(s)} = g_{\boldsymbol{\Phi}}\PC{g_{\boldsymbol{\varphi}}\PC{\bmf_n^{(s)}}} \in \opR^{D''}$.
We implement both $g_{\boldsymbol{\varphi}}$ and $g_{\boldsymbol{\Phi}}$ as MLPs with weights and biases described by $\boldsymbol{\varphi}$ and $\boldsymbol{\Phi}$, respectively. 
We learn the AE function $g_{\boldsymbol{\Phi}}\PC{g_{\boldsymbol{\varphi}}\PC{\cdot}}$ with an MSE loss given by
\begin{align}\label{eq:loss_mse_autoencoder}
    \loss_{\text{AE}} = \frac{1}{|\setN_s|} \sum_{n\in\setN_s}  \vecnorm{\bmf_n^{(s)}-g_{\boldsymbol{\Phi}}\PC{g_{\boldsymbol{\varphi}}\PC{\bmf_n^{(s)}}}}^2,
\end{align}
where $\bmf_n^{(s)}$ is given in \fref{eq:feature_autoencoder}.
This loss function is self-supervised and solely relies on CSI. 

We implement {the AE encoding function}, $g_{\boldsymbol{\varphi}}$, as an MLP with $\{D'',32,D'\}$ activations per layer and the AE decoding function, $g_{\boldsymbol{\Phi}}$, as an MLP with $\{D',32,D''\}$ activations per layer. 
According to our experiments, this MLP architecture has been proven to be sufficient for accurate CSI reconstruction, considering an encoding dimension of $D'=64$ and adapting the size of our hidden layer from $32$ to $128$ activations.

We implement the scenario-specific POS function, $p_{\boldsymbol{\theta}_s}$ (cf. \fref{sec:positioning}) and the scenario-specific CC function $p_{\boldsymbol{\theta}_s}$ (cf. \fref{sec:cc}), for every scenario $s$ as an MLP with $\{D',12,8,6,4,d\}$ activations per layer, except in the case of CC-PCA, which we detail in \fref{sec:pca}.
For all the MLPs, all layers except for the last one have ReLU activations. We initialize all weights using Glorot initialization \cite{Glorot2010} and train the network with Adam.

\subsubsection{Autoencoder With Augmentation (AE-AUG)}
As our third baseline method, we use the same AE as in the previous section, but now with AUG. 
The goal of this baseline is to evaluate the POS and CC performance of AE architectures commonly used for learning vector embeddings \cite{Alikhani2024,Ott2024,Aboulfotouh2025,Guler2025}---now considering diverse RSs---but not specifically tailored to POS and CC applications. For this method, we assume the exact same procedure as described in \fref{sec:ae_without_aug}. The key difference now is that $\bA_n^{(s)} \neq \underline{\hat\bH}_n^{(s)}$.

\subsection{Performance Metrics} \label{sec:performance_metrics}
We evaluate the effectiveness of the proposed methods and baselines using two metrics related to POS (cf. \fref{sec:positioning}) and four metrics related to CC (cf. \fref{sec:cc}). 
For brevity, we refer to \cite[Sec.~IV-C]{Taner2024} for the details.  
(i) \emph{Mean distance error (MDE)} represents the average Euclidean norm error in the UE position estimates across all UE positions.
(ii) \emph{95th percentile distance error} denotes the 95th percentile of the Euclidean norm error in the UE position estimates across all UE positions. 
(iii) \emph{Trustworthiness (TW)} penalizes neighborhood relationships in latent space that are absent at ground-truth UE positions; 
TW values range from $[0,1]$, with optimal value~$1$. 
(iv) \emph{Continuity (CT)} quantifies how well neighborhood relationships in the ground-truth UE positions are preserved in the latent space; CT values range from $[0,1]$, with optimal value~$1$.
(v) \emph{Kruskal stress (KS)} measures the discrepancy between pairwise distances in the ground-truth UE positions and their corresponding distances in the latent space; KS values range from $[0,1]$, with optimal value~$0$.
(vi) \emph{Rajski distance (RD)} evaluates the difference between the mutual information and joint entropy of the distribution of pairwise distances in the ground-truth UE positions and latent space; RD values range from $[0,1]$, with optimal value~$0$. 
%


\section{Results}
\label{sec:results}

\subsection{Simulation Settings}

We now evaluate the efficacy of the proposed CSI2Vec methods and baselines across $S=3$ distinct scenarios: a simulated outdoor scenario, a simulated indoor scenario, and a measurement-based indoor scenario. 
\fref{tbl:simulation_parameters} summarizes the parameters of each scenario.
Following \fref{eq:max_values}, we use $A_{\text{max}} = 8$, $B_{\text{max}} = 8$, $U_{\text{max}} = 1$, and $W_{\text{max}} = 1200$. 
Thus, $\underline{\hat\bH}_n^{(s)} \in \opC^{8 \times 8 \times 1 \times 1200}$, where $n\in \setN_s$ and $s \in \setS$.

\begin{table*}
\centering
\caption{Parameters for each evaluated Scenario.}\label{tbl:simulation_parameters}
\begin{tabular}{@{}llll@{}}
\toprule
 & \multicolumn{3}{c}{Value or type of parameter for each scenario} \\
\cmidrule(){2-4}
Parameter & Simulated outdoor \cite{remcom} & Simulated indoor \cite{remcom} & Measured indoor~\cite{dataset-dichasus-cf0x} \\
\midrule
Number of APs & $A_1=6$  & $A_2=8$ & $A_3=4$ \\
Number of antennas per AP & $B_1=4$ & $B_2=4$  & $B_3=8$  \\
APs' antenna array structure & Uniform linear array & Uniform linear array  & Uniform rectangular array, $2\times 4$  \\
AP antenna height & $10$\,m & $2$\,m  & $1.56$\,m-$1.78$\,m \\
AP antenna spacing & Half-wavelength &  Half-wavelength  &  Half-wavelength  \\
Number of UE positions & $N_1=14\,642$ & $N_2=14\,606$ & $N_3=83\,985$ \\
Spacing between UE positions &40\,cm & 2.5\,cm & (not fixed) \\
Number of antennas per UE & $U_1=1$ & $U_2=1$ & $U_3=1$ \\
UE antenna height & $1.5$\,\text{m} & $1.5$\,\text{m} & $\approx 0.94\,\text{m}$ \\
AP and UE antenna type & Omnidirectional &  Omnidirectional & Omnidirectional \\
Carrier frequency & $1.9$\,GHz & $2.4$\,GHz & $1.272$\,GHz \\
Bandwidth & $20$\,MHz & $20$\,MHz & $50$\,MHz \\
Number of used subcarriers & $W_1=1200$ & $W_2=64$ & $W_3=1024$\\
\bottomrule
\end{tabular}
\end{table*}

In the following experiments, we consider $q=20\%$ removed subcarriers. 
For the simulated outdoor and indoor scenarios, we add {i.i.d. circularly-symmetric complex Gaussian} noise with variance $N_0$, as described in \fref{sec:adding_noise}, selecting uniformly at random the highest SNR per AP to be within the range of $[10,21]$\,dB.
For the measurement-based indoor scenario,  the range is $[25,41]$\,dB.\footnote{We choose lower SNR ranges for simulation-based scenarios compared to measurement-based scenarios because measurements already include channel estimation errors, whereas simulations do not.}
We truncate the maximum number of channel taps to $C^{'}_{\text{max}}=16$ taps.
Therefore, the input feature dimension $D$ is given by {$D = A_{\text{max}} B_{\text{max}} U_{\text{max}} C^{'}_{\text{max}} = 1024$}. For both baselines AE-AUG and AE, we set {$D''= 2A_{\text{max}} B_{\text{max}} U_{\text{max}} C^{'}_{\text{max}} = 2048$}. 
Unless stated otherwise, we set our vector embedding dimension to $D'=16$.

For training and testing, we use $240$ and $120$ batches, respectively, with a batch size of $100$ samples.
To ensure a balanced representation across datasets, each batch is constructed by first selecting a scenario uniformly at random, followed by selecting a sample from that scenario uniformly at random. When triplets are used, we take the sample as the anchor and we pick a triplet uniformly at random.
All results and performance metrics are computed using the test set.

In the following subsections, we provide a detailed description of each scenario and discuss the corresponding results. 
These include plots of ground-truth and estimated UE positions, as well as SN-based channel charts. 
We then evaluate the performance metrics defined in~\fref{sec:performance_metrics} for both the proposed methods and baseline approaches for the downstream tasks of POS, CC-SN, and CC-PCA (cf. \fref{sec:downstream_tasks}), where we set $d=2$.

\begin{figure*}[h]
    \centering
    \hfill
    \subfigure[Simulated outdoor scenario]
    {
    \label{fig:gt_outdoor_scenario}
    \includegraphics[width=0.33\textwidth,height=6.5cm]{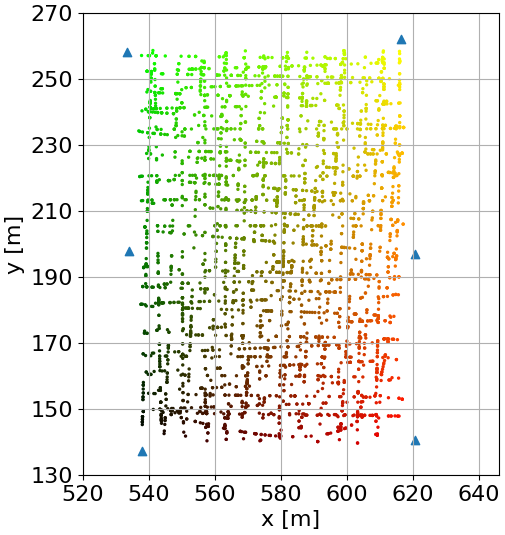}
    }
    \hfill
    \subfigure[Simulated indoor scenario]  
    {
    \label{fig:gt_indoor_scenario}
    \includegraphics[width=0.25\textwidth,height=6.5cm]{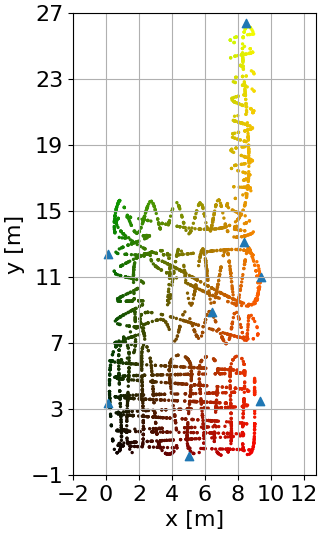}
    }
    \hfill
    \subfigure[Measurement-based indoor scenario]
    {
    \label{fig:gt_dichasus_scenario}
    \includegraphics[width=0.33\textwidth,height=6.5cm]{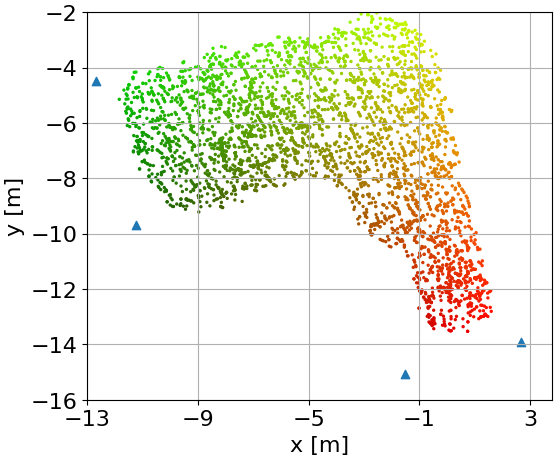}
    } \hspace{40pt}
    \hfill
    
    \caption{Ground-truth UE positions for the three presented scenarios (green-to-red gradient colored area), AP positions (blue triangles): (a) simulated outdoor scenario, (b) simulated indoor scenario, (c) measurement-based indoor scenario.} 
    \label{fig:gt_all_scenarios}  
\end{figure*}

\fref{fig:gt_all_scenarios} shows the ground-truth UE positions for each of the scenarios. 
Figs.~\ref{fig:outdoor_results}, \ref{fig:indoor_results}, and \ref{fig:dichasus_results} visually compare the estimated UE positions generated through POS and the channel charts generated through CC-SN for each scenario. 
As in \cite{Studer2018}, we normalize the channel charts generated through CC-SN.
Tbls.~\ref{tbl:simulated_outdoor}, \ref{tbl:simulated_indoor}, and~\ref{tbl:measured_indoor} provide the respective performance metrics.
When evaluating the TW and CT metrics, we set $J=0.05N$ as in~\cite{Studer2018}; when evaluating the RD metric, we quantize the pairwise distances into $20$ uniform bins \cite{Taner2024}.

CC-SN and CC-PCA generate channel charts in arbitrary coordinates, which results in (i) unitless axes that are not measured in meters and (ii) distance error metrics becoming irrelevant in the performance analysis.
Therefore, we exclude these metrics for all proposed methods and baselines in Tables~\ref{tbl:simulated_outdoor}, \ref{tbl:simulated_indoor}, and~\ref{tbl:measured_indoor}. 
Additionally, to avoid redundancy, we omit the visualizations of estimated UE positions and channel charts for the AE baseline, as they closely resemble those of the AE-AUG. 
However, we still include its performance metrics in Tbls.~\ref{tbl:simulated_outdoor}, \ref{tbl:simulated_indoor}, and~\ref{tbl:measured_indoor} for completeness.

For the proposed method CSI2Vec-AUG-SEMI (cf. \fref{sec:csi2vec_SEMI}), we assume that $\setS' = \{2\}$, i.e., we have ground-truth UE positions only for the simulated indoor scenario. 

\subsection{Simulated Outdoor Scenario} \label{sec:simulated_outdoor}
\subsubsection{Description}
In this evaluation, we use channel vectors from Remcom's Wireless InSite ray-tracing software \cite{remcom}.
The considered scenario represents an urban environment in which six APs are distributed around a rectangular area of size $83\,\text{m} \times 122\,\text{m}$. 
The UE follows a meandering trajectory in both north-south and east-west directions to cover the entire region; \fref{fig:gt_outdoor_scenario} illustrates the UE and AP positions. For this scenario, we set $T_c = 10$ in \fref{eq:triplets_same_scenario} and \fref{eq:triplets_diff_scenario}. 
The first column of \fref{tbl:simulation_parameters} provides an overview of the used simulation parameters. 
{According to \fref{tbl:simulation_parameters}}, {$\hat\bH^{(1)}_{n} \in \opC^{6\times4\times1\times1200}$}, where $n\in\setN_1$.

\subsubsection{Performance Evaluation}
\begin{figure*}[h]
    \centering
    \setlength{\tabcolsep}{0pt} 
    \renewcommand{\arraystretch}{0}   
     \subfigure[CSI2Vec]
    {
    \label{fig:outdoor_csi2vec_pos}
    \resizebox{0.18\textwidth}{3.35cm}{\includegraphics{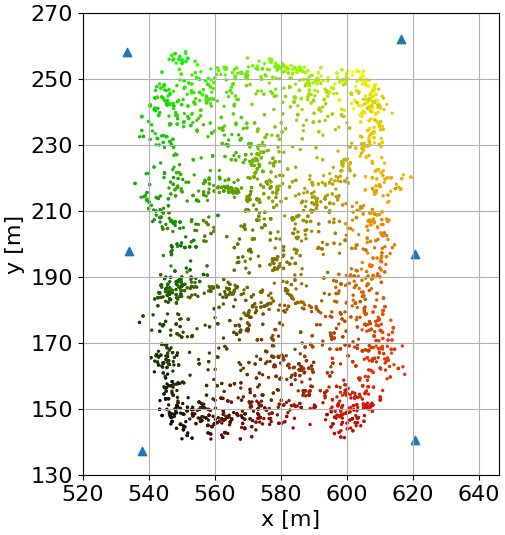}
    }}\hspace{-3pt}
    \subfigure[CSI2Vec-AUG]
    {
    \label{fig:outdoor_csi2vec_aug_pos}
    \resizebox{0.18\textwidth}{3.35cm}{\includegraphics{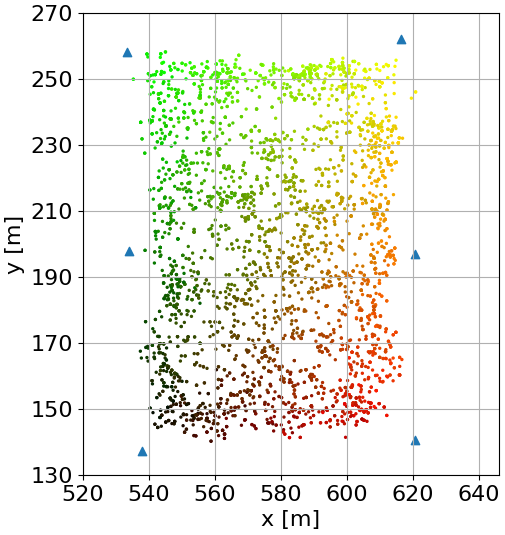}
    }}\hspace{-3pt}
    \subfigure[CSI2Vec-AUG-SS]
    {
    \label{fig:outdoor_csi2vec_aug_ss_pos}
    \resizebox{0.18\textwidth}{3.35cm}{\includegraphics{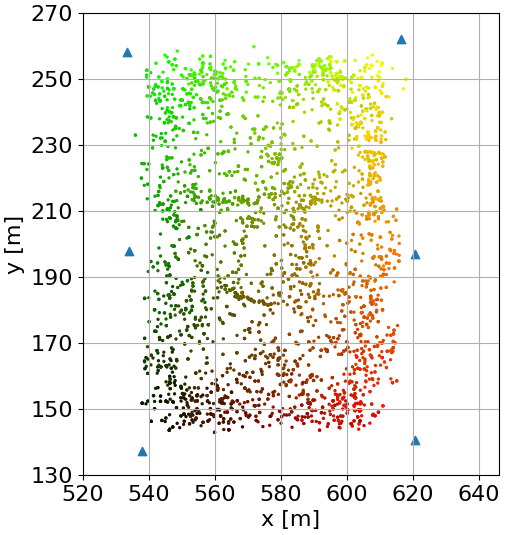}
    }}
     \subfigure[SCS-EE]  
    {
    \label{fig:outdoor_scs_ee_pos}
    \resizebox{0.18\textwidth}{3.35cm}{\includegraphics{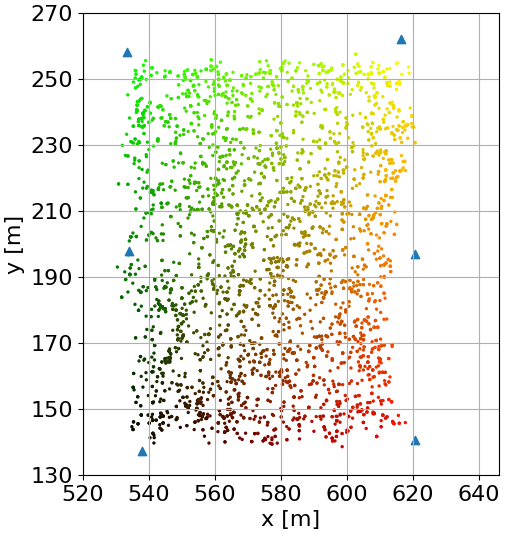}
    }}\hspace{-3pt}
    \subfigure[AE-AUG]
    {
    \label{fig:outdoor_ae_aug_pos}
    \resizebox{0.18\textwidth}{3.35cm}{\includegraphics{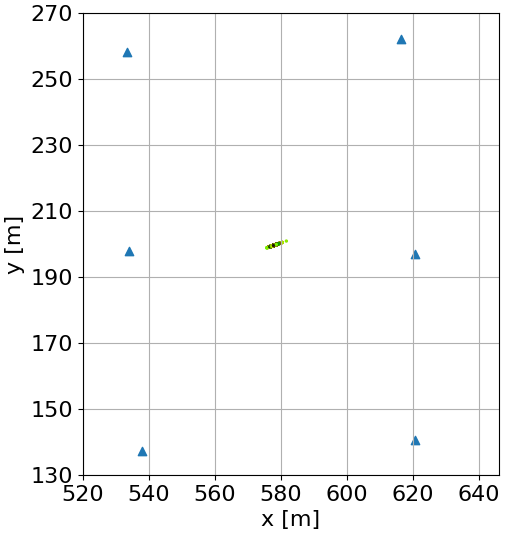}
    }}\hspace{-3pt}
     \subfigure[CSI2Vec] 
    {
    \label{fig:outdoor_csi2vec_cc}
    \resizebox{0.18\textwidth}{3.35cm}
    {\includegraphics{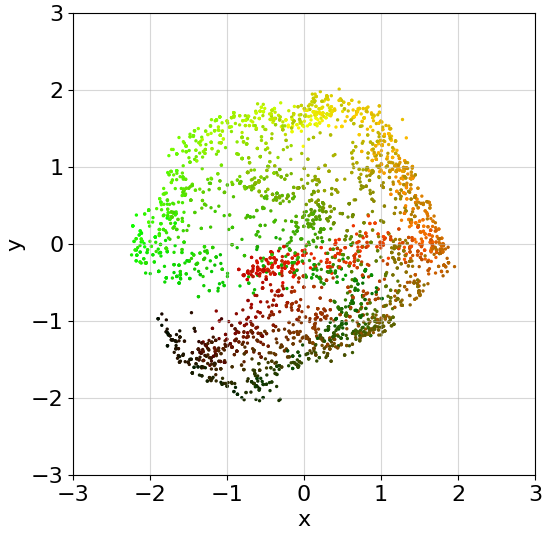}
    }}\hspace{-3pt}
    \subfigure[CSI2Vec-AUG]
    {
    \label{fig:outdoor_csi2vec_aug_cc}
    \resizebox{0.18\textwidth}{3.35cm}
    {\includegraphics{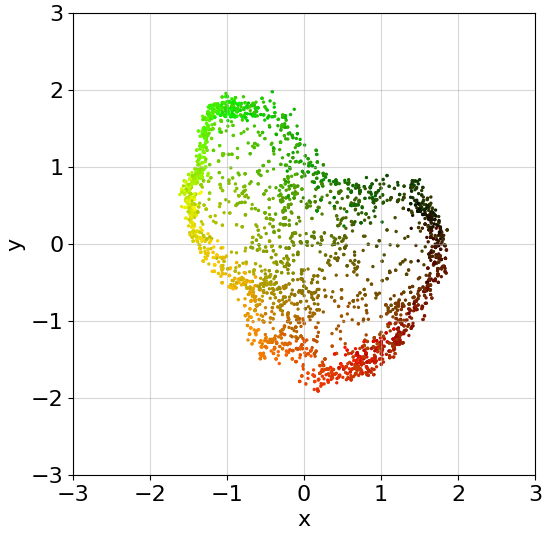}
    }}\hspace{-3pt}
    \subfigure[CSI2Vec-AUG-SEMI]
    {
    \label{fig:outdoor_csi2vec_aug_ss_cc}
    \resizebox{0.18\textwidth}{3.35cm}
    {\includegraphics{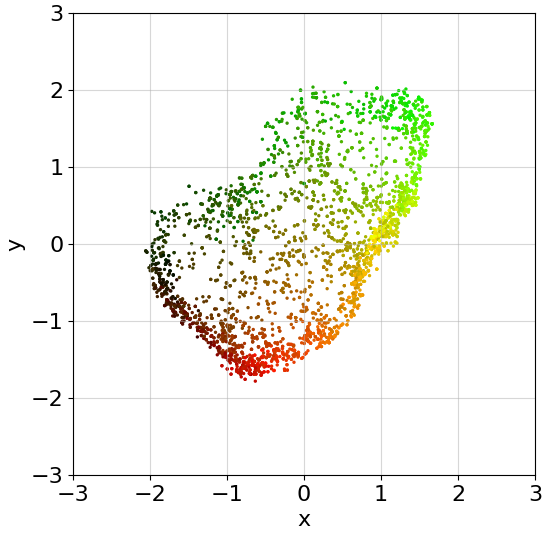}
    }}
    \subfigure[SCS-EE]  
    {
    \label{fig:outdoor_scs_ee_cc}
    \resizebox{0.18\textwidth}{3.35cm}
    {\includegraphics{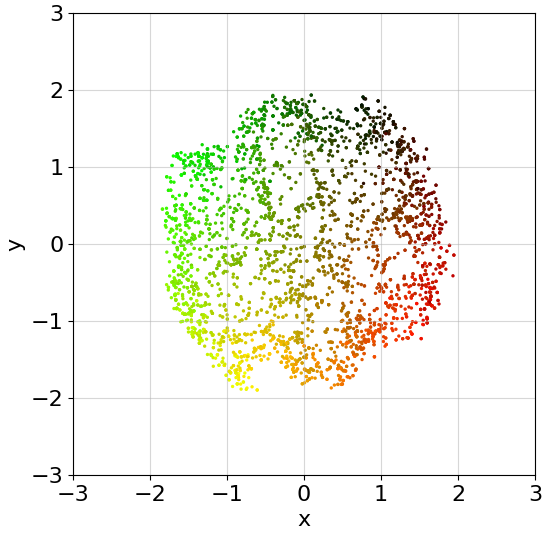}
    }}\hspace{-3pt}
    \subfigure[AE-AUG]
    {
    \label{fig:outdoor_ae_aug_cc}
    \resizebox{0.18\textwidth}{3.35cm}
    {\includegraphics{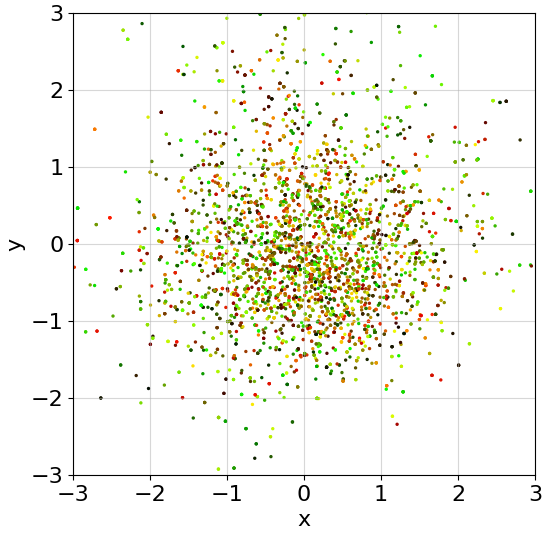}
    }}\hspace{-3pt}
    \caption{Results for the simulated outdoor scenario: (a-e) POS estimates and (f-j) channel charts using CC-SN for both the proposed and baseline methods. 
    The proposed CSI2Vec-AUG method achieves an MDE, which is only $0.898$\,m higher than that of the SCS-EE baseline for POS, and delivers comparable CC performance to SCS-EE for CC-SN, despite using a $64\times$ smaller input feature dimension. 
    The AE-AUG baseline fails to generate a coherent estimated POS map and channel chart.
    } 
    \label{fig:outdoor_results}  
\end{figure*}

In Figs. \ref{fig:outdoor_csi2vec_pos}-(e), we present the estimated POS maps for the proposed methods and baselines. 
As observed, all methods---except for AE-AUG---produce rectangular POS maps that preserve the color gradient and closely resemble the ground-truth UE positions. \fref{tbl:simulated_outdoor} shows that CSI2Vec-AUG achieves an {MDE which is only $0.898$\,m higher than that of the} SCS-EE {baseline}, despite using significantly smaller input features (i.e., only $D'=16$ instead of $1024$).
Additionally, AUG improves POS performance, whereas CSI2Vec with SEMI does not provide an advantage, as the ground-truth UE positions do not belong to the evaluated scenario. 
Examining the estimated POS map of AE-AUG, \fref{fig:outdoor_ae_aug_pos}, we observe that despite having ground-truth UE positions for all samples, the vector embeddings produced by the AE-AUG’s encoding function fail to capture spatial relationships, leading to poor POS performance.  
We reach the same conclusion for the AE.

In Figs. \ref{fig:outdoor_csi2vec_cc}-(j) and \fref{tbl:simulated_outdoor}, we present the channel charts for CC-SN and the performance metrics for CC-SN and CC-PCA, respectively, for the proposed methods and baselines. 
As observed, all methods—except for AE-AUG—produce channel charts that preserve the color gradient of the ground-truth UE positions.
The results indicate that the CSI2Vec-based methods generate channel charts comparable in quality to those obtained through SCS-EE, despite using significantly smaller input features (i.e., only $D'=16$ instead of  $1024$).  
Examining the performance metrics of AE-AUG and AE, we observe that their vector embeddings fail entirely to capture spatial relationships, resulting in poor channel charts.

\begin{table*}
\centering
\hspace{-0.25cm}
\resizebox{0.99\textwidth}{!}{
\begin{minipage}[c]{0.99 \textwidth}
    \centering
    \caption{POS and CC performance comparison for the simulated outdoor scenario.} 
    \label{tbl:simulated_outdoor}
    \begin{tabular}{@{}lccccccccc@{}}
        \toprule
        & & & & \multicolumn{2}{c}{Positioning error [m]} & \multicolumn{4}{c}{Latent space quality metrics} \\
        \cmidrule(lr){5-6} \cmidrule(lr){7-10}  
        Method &  {Downstream task} & Figure & Dim. & Mean$\,\downarrow$ & 95th percentile$\,\downarrow$ & TW$\,\uparrow$ & CT$\,\uparrow$ & KS$\,\downarrow$ & RD$\,\downarrow$  \\
        \midrule
        \multirow{3}{*}{CSI2Vec (across scenarios)} 
        & POS & \fref{fig:outdoor_csi2vec_pos} & 16 & 4.73 & 10.275 & 0.991 & 0.993 & 0.095 & 0.649 \\ 
        & CC-SN & \fref{fig:outdoor_csi2vec_cc} & 16 & -- & -- & 0.908 & 0.976 & 0.354 & 0.903 \\
        & CC-PCA & -- & 16 & -- & -- & 0.901 & 0.975 & 0.355 & 0.903 \\
        \midrule
        \multirow{3}{*}{CSI2Vec-AUG (across scenarios)} 
        & POS & \fref{fig:outdoor_csi2vec_aug_pos} & 16 & 4.026 & 8.639 & 0.993 & 0.994 & 0.083 & 0.612 \\ 
        & CC-SN & \fref{fig:outdoor_csi2vec_aug_cc} & 16 & -- & -- & 0.981 & 0.988 & 0.165 & 0.783\\
        & CC-PCA & -- & 16 & -- & -- & 0.980 & 0.987 & 0.169 & 0.788\\
        \midrule
        \multirow{3}{*}{CSI2Vec-AUG-SEMI (across scenarios)} 
        & POS & \fref{fig:outdoor_csi2vec_aug_ss_pos} & 16 & 4.523 & 9.778 & 0.992 & 0.993 & 0.090 & 0.636 \\ 
        & CC-SN & \fref{fig:outdoor_csi2vec_aug_ss_cc} & 16 & -- & -- & 0.980 & 0.987 & 0.214 & 0.823 \\
        & CC-PCA & -- & 16 & -- & -- & 0.970 & 0.982 & 0.236 & 0.843\\
        \midrule
        \multirow{3}{*}{SCS-EE} 
        & POS & \fref{fig:outdoor_scs_ee_pos} & 1024 & 3.128 & 6.199 & 0.999 & 0.998 & 0.042 & 0.449 \\ 
        & CC-SN & \fref{fig:outdoor_scs_ee_cc} & 1024 & -- & -- & 0.991 & 0.991 & 0.163 & 0.776 \\
        & CC-PCA & -- & 1024 & -- & -- & 0.964 & 0.981 & 0.234 & 0.859\\
        \midrule
        \multirow{3}{*}{AE (across scenarios)} 
        & POS & -- & 16 & 38.321 & 60.095 & 0.525 & 0.524 & 0.702 & 0.998  \\ 
        & CC-SN & -- & 16 & -- & -- & 0.539 & 0.535 & 0.598 & 0.996 \\
        & CC-PCA & -- & 16 & -- & -- & 0.536 & 0.534 & 0.605 & 0.997 \\
        \midrule
        \multirow{3}{*}{AE-AUG (across scenarios)} 
        & POS & \fref{fig:outdoor_ae_aug_pos} & 16 & 38.340 & 60.165 & 0.518 & 0.524 & 0.700 & 0.999 \\ 
        & CC-SN & \fref{fig:outdoor_ae_aug_cc} & 16 & -- & -- & 0.523 & 0.526 & 0.610 & 0.999\\
        & CC-PCA & -- & 16 & -- & -- & 0.523 & 0.528 & 0.633 & 0.999 \\
        \bottomrule
    \end{tabular}		
\end{minipage}}
\end{table*}

\subsection{Simulated Indoor Scenario} \label{sec:simulated_indoor}

\subsubsection{Description}

In this evaluation, we use channel vectors from Remcom's Wireless InSite ray-tracing software \cite{remcom}. This scenario represents an indoor office environment in which eight APs are distributed in an area. The UE follows a meandering trajectory to cover the entire office space; \fref{fig:gt_indoor_scenario} illustrates the UE and AP positions. For this scenario, we set $T_c = 1$ in \fref{eq:triplets_same_scenario} and \fref{eq:triplets_diff_scenario}. The second column of \fref{tbl:simulation_parameters} provides an overview of the simulation parameters. 
According to \fref{tbl:simulation_parameters}, {$\hat\bH^{(2)}_{n} \in \opC^{8\times4\times1\times64}$}, where $n\in\setN_2$.

\subsubsection{Performance Evaluation}

\begin{figure*}[h]
    \centering
     \subfigure[CSI2Vec]
    {
    \label{fig:indoor_csi2vec_pos}
    \resizebox{0.18\linewidth}{4.5cm}{\includegraphics{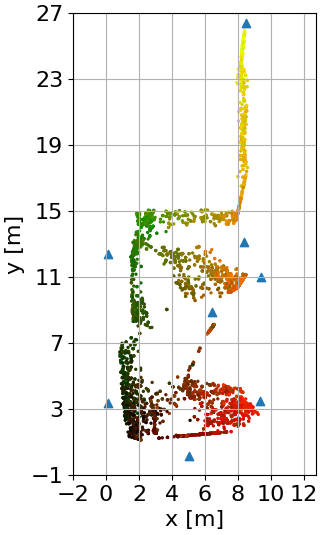}
    }}\hspace{-4pt}
    \subfigure[CSI2Vec-AUG]
    {
    \label{fig:indoor_csi2vec_aug_pos}
    \resizebox{0.18\linewidth}{4.5cm}{\includegraphics{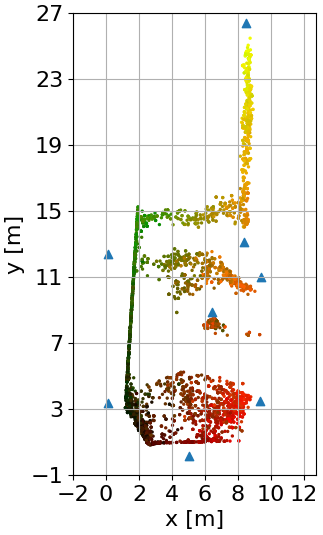}
    }}\hspace{-4pt}
    \subfigure[CSI2Vec-AUG-SEMI]
    {
    \label{fig:indoor_csi2vec_aug_ss_pos}
    \resizebox{0.18\linewidth}{4.5cm}{\includegraphics{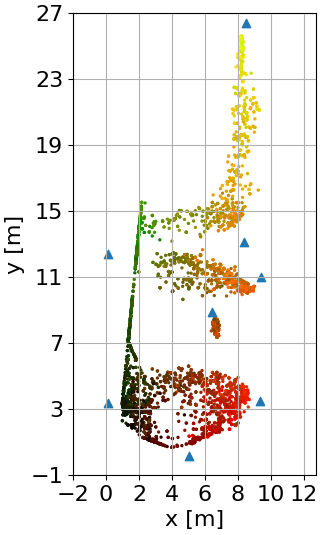}
    }}
    \subfigure[SCS-EE]  
    {
    \label{fig:indoor_scs_ee_pos}
    \resizebox{0.18\textwidth}{4.5cm}{\includegraphics{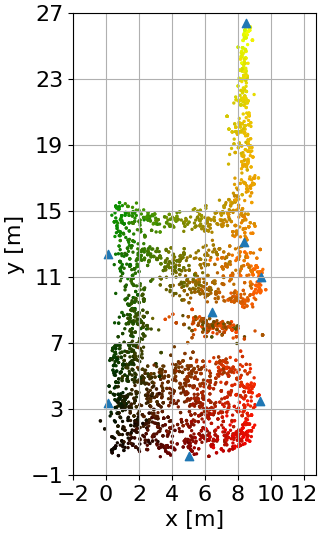}
    }}\hspace{-4pt}
    \subfigure[AE-AUG]
    {
    \label{fig:indoor_ae_aug_pos}
    \resizebox{0.18\linewidth}{4.5cm}{\includegraphics{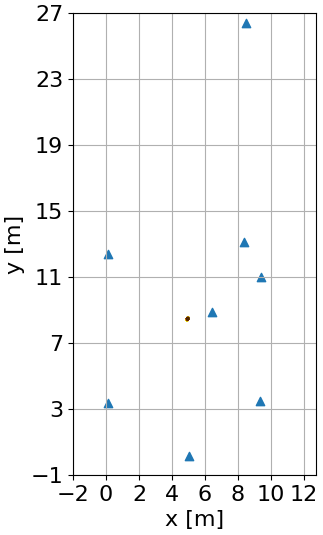}
    }}\hspace{-4pt}
     \subfigure[CSI2Vec] 
    {
    \label{fig:indoor_csi2vec_cc}
    \resizebox{0.18\linewidth}{3.35cm}
    {\includegraphics{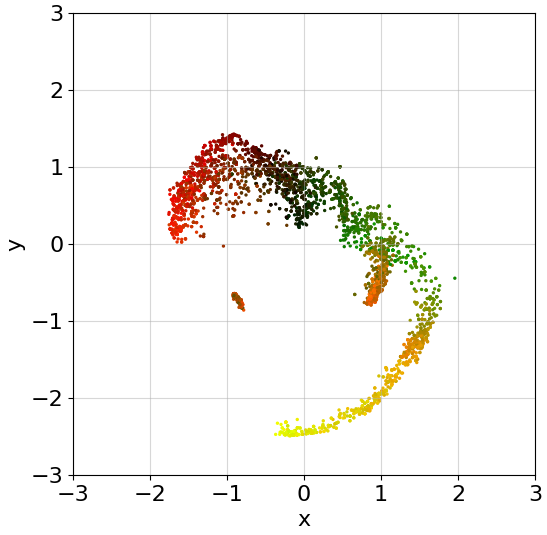}
    }}\hspace{-4pt}
    \subfigure[CSI2Vec-AUG]
    {
    \label{fig:indoor_csi2vec_aug_cc}
    \resizebox{0.18\linewidth}{3.35cm}
    {\includegraphics{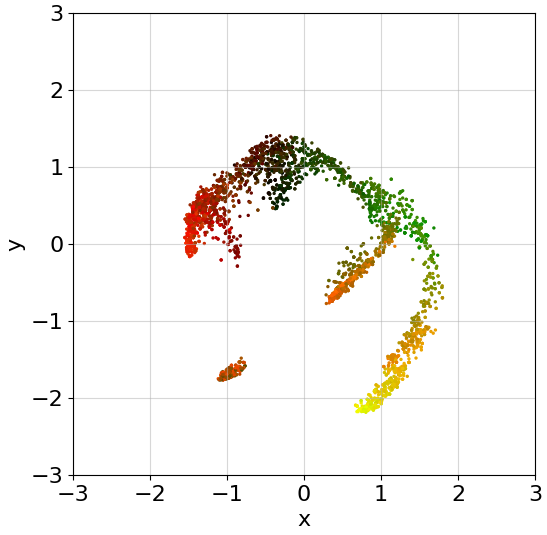}
    }}\hspace{-4pt}
    \subfigure[CSI2Vec-AUG-SEMI]
    {
    \label{fig:indoor_csi2vec_aug_ss_cc}
    \resizebox{0.18\linewidth}{3.35cm}
    {\includegraphics{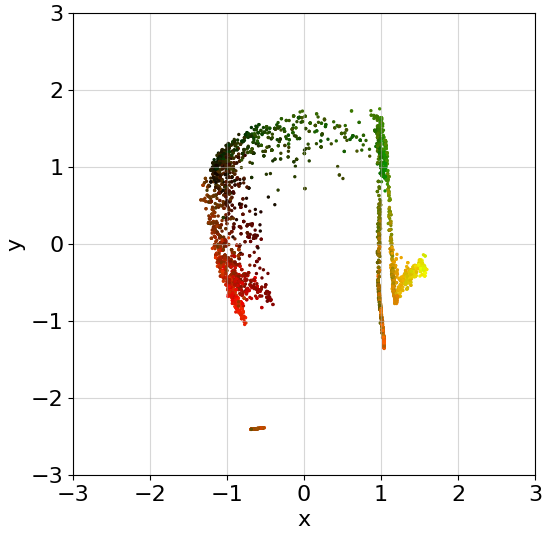}
    }}
    \subfigure[SCS-EE]  
    {
    \label{fig:indoor_scs_ee_cc}
    \resizebox{0.18\linewidth}{3.35cm}
    {\includegraphics{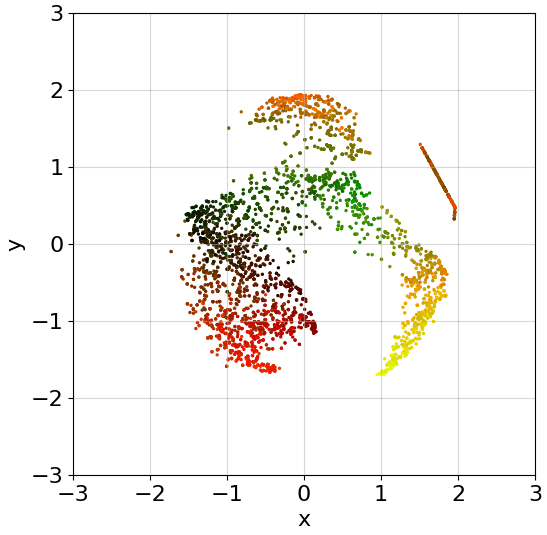}
    }}\hspace{-4pt}
    \subfigure[AE-AUG]
    {
    \label{fig:indoor_ae_aug_cc}
    \resizebox{0.18\linewidth}{3.35cm}
    {\includegraphics{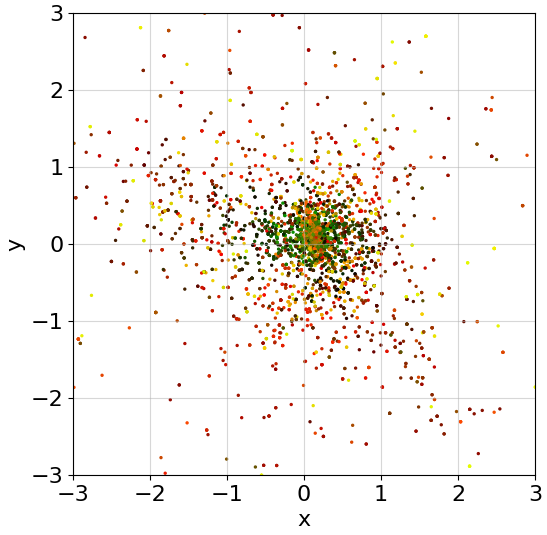}
    }}\hspace{-4pt}
    \caption{Results for the simulated indoor scenario: (a-e) POS estimates and (f-j) channel charts using CC-SN for both the proposed and baseline methods. 
    The proposed CSI2Vec-AUG method achieves an MDE, which is only $0.493$\,m higher than that of the SCS-EE baseline for POS, and delivers comparable CC performance to SCS-EE for CC-SN, despite using a $64\times$ smaller input feature dimension. 
    The AE-AUG baseline fails to generate a coherent estimated POS map and channel chart.
    } 
    \label{fig:indoor_results}  
\end{figure*}

In Figs. \ref{fig:indoor_csi2vec_pos}-(e), we present the estimated POS maps for both the proposed methods and baselines.
As observed, all methods---except for AE-AUG---produce well-shaped POS maps that preserve the color gradient and closely {resemble} the ground-truth UE positions, as expected. \fref{tbl:simulated_indoor} shows that CSI2Vec-AUG achieves an {MDE which is only $0.493$\,m higher than that of the} SCS-EE {baseline}, despite using significantly smaller input features 
(i.e., only $D'=16$ instead of  $1024$).
Additionally, AUG and SEMI improve POS performance, as the ground-truth UE positions belong to the evaluated scenario. 
Examining the estimated POS map of AE-AUG, \fref{fig:indoor_ae_aug_pos}, we observe that despite having ground-truth UE positions for all samples, the vector embeddings produced by the AE-AUG’s encoding function fail to capture spatial relationships, leading to poor POS performance. 
We reach the same conclusion for the AE.

In Figs. \ref{fig:indoor_csi2vec_cc}-(j) and \fref{tbl:simulated_indoor}, we present the channel charts for CC-SN and the performance metrics for CC-SN and CC-PCA, respectively, for both the proposed methods and baselines. 
As observed, all methods---except for AE-AUG---produce channel charts that preserve the color gradient of the ground-truth UE positions. 
The results indicate that the CSI2Vec-based methods generate channel charts comparable in quality to those obtained through SCS-EE, despite using significantly smaller input features (i.e., only $D'=16$ instead of  $1024$).
Examining the performance metrics of AE-AUG and AE, we observe that their vector embeddings fail entirely to capture spatial relationships, resulting in poor channel charts.

\begin{table*}
\centering
\hspace{-0.25cm}
\resizebox{0.99\textwidth}{!}{
\begin{minipage}[c]{0.99 \textwidth}
    \centering
    \caption{POS and CC performance comparison for the simulated indoor scenario.} 
    \label{tbl:simulated_indoor}
    \begin{tabular}{@{}lccccccccc@{}}
        \toprule
        & & & & \multicolumn{2}{c}{Positioning error [m]} & \multicolumn{4}{c}{Latent space quality metrics} \\
        \cmidrule(lr){5-6} \cmidrule(lr){7-10}  
        Method &  {Downstream task} & Figure & Dim. & Mean$\,\downarrow$ & 95th percentile$\,\downarrow$ & TW$\,\uparrow$ & CT$\,\uparrow$ & KS$\,\downarrow$ & RD$\,\downarrow$  \\
        \midrule
        \multirow{3}{*}{CSI2Vec (across scenarios)} 
        & POS & \fref{fig:indoor_csi2vec_pos} & 16 & 1.202 & 2.528 & 0.960 & 0.963 & 0.154 & 0.734 \\ 
        & CC-SN & \fref{fig:indoor_csi2vec_cc} & 16 & -- & -- & 0.948 & 0.948 & 0.299 & 0.873 \\
        & CC-PCA & -- & 16 & -- & -- & 0.938 & 0.940 & 0.304 & 0.871 \\
        \midrule
        \multirow{3}{*}{CSI2Vec-AUG (across scenarios)} 
        & POS & \fref{fig:indoor_csi2vec_aug_pos} & 16 & 1.164 & 2.599 & 0.963 & 0.966 & 0.150 & 0.729 \\ 
        & CC-SN & \fref{fig:indoor_csi2vec_aug_cc} & 16 & -- & -- & 0.944 & 0.946 & 0.332 & 0.884 \\
        & CC-PCA & -- & 16 & -- & -- & 0.940 & 0.941 & 0.334 & 0.882\\
        \midrule
        \multirow{3}{*}{CSI2Vec-AUG-SEMI (across scenarios)} 
        & POS & \fref{fig:indoor_csi2vec_aug_ss_pos} & 16 & 1.125 & 2.396 & 0.965 & 0.967 & 0.144 & 0.719 \\ 
        & CC-SN & \fref{fig:indoor_csi2vec_aug_ss_cc} & 16 & -- & -- & 0.948 & 0.937 & 0.392 & 0.898 \\
        & CC-PCA & -- & 16 & -- & -- & 0.935 & 0.928 & 0.430 & 0.918\\
        \midrule
        \multirow{3}{*}{SCS-EE} 
        & POS & \fref{fig:indoor_scs_ee_pos} & 1024 & 0.671 & 1.651 & 0.985 & 0.986 & 0.089 & 0.59 \\ 
        & CC-SN & \fref{fig:indoor_scs_ee_cc} & 1024 & -- & -- & 0.948 & 0.934 & 0.462 & 0.924 \\
        & CC-PCA & -- & 1024 & -- & -- & 0.941 & 0.941 & 0.396 & 0.907 \\
        \midrule
        \multirow{3}{*}{AE (across scenarios)} 
        & POS & -- & 16 & 5.884 & 13.122 & 0.568 & 0.570 & 0.740 & 0.982 \\ 
        & CC-SN & -- & 16 & -- & -- & 0.548 & 0.565 & 0.746 & 0.993 \\
        & CC-PCA & -- & 16 & -- & -- & 0.580 & 0.576 & 0.775 & 0.989 \\
        \midrule
        \multirow{3}{*}{AE-AUG (across scenarios)} 
        & POS & \fref{fig:indoor_ae_aug_pos} & 16 & 5.884 & 13.130 & 0.592 & 0.598 & 0.794 & 0.989 \\ 
        & CC-SN & \fref{fig:indoor_ae_aug_cc} & 16 & -- & -- & 0.627 & 0.591 & 0.743 & 0.993 \\
        & CC-PCA & -- & 16 & -- & -- & 0.657 & 0.593 & 0.780 & 0.995\\
        \bottomrule
    \end{tabular}		
\end{minipage}}
\end{table*}

\subsection{Measurement-Based Indoor Scenario} \label{sec:measured_indoor}

\subsubsection{Description}
In this evaluation, we use the channel vectors from the measured DICHASUS dataset \cite{dataset-dichasus-cf0x}. 
This scenario corresponds to an indoor factory environment in which four APs are distributed in an area. 
The UE follows a meandering trajectory to cover the entire factory space; \fref{fig:gt_dichasus_scenario} illustrates the UE and AP positions. 
For this scenario, we set $T_c = 1$ in \fref{eq:triplets_same_scenario} and \fref{eq:triplets_diff_scenario}. 
The third column of \fref{tbl:simulation_parameters} provides an overview of the simulation parameters.
According to \fref{tbl:simulation_parameters}, {$\hat\bH^{(3)}_{n} \in \opC^{4\times8\times1\times1024}$}, where $n\in\setN_3$.

\subsubsection{Performance Evaluation}
\begin{figure*}[h]
    \centering
     \subfigure[CSI2Vec]
    {
    \label{fig:dichasus_csi2vec_pos}
    \resizebox{0.18\textwidth}{3.35cm}{\includegraphics{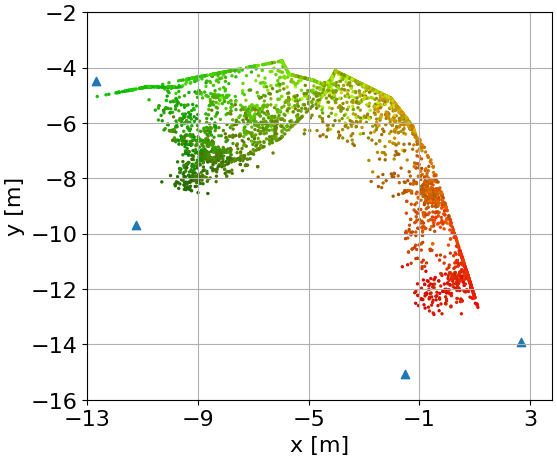}
    }}\hspace{-4pt}
    \subfigure[CSI2Vec-AUG]
    {
    \label{fig:dichasus_csi2vec_aug_pos}
    \resizebox{0.18\textwidth}{3.35cm}{\includegraphics{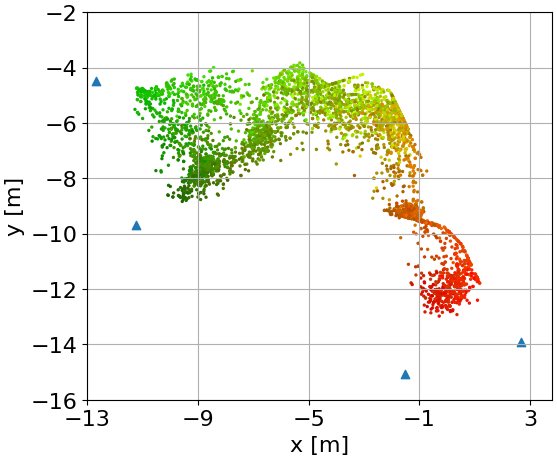}
    }}\hspace{-4pt}
    \subfigure[CSI2Vec-AUG-SEMI]
    {
    \label{fig:dichasus_csi2vec_aug_ss_pos}
    \resizebox{0.18\textwidth}{3.35cm}{\includegraphics{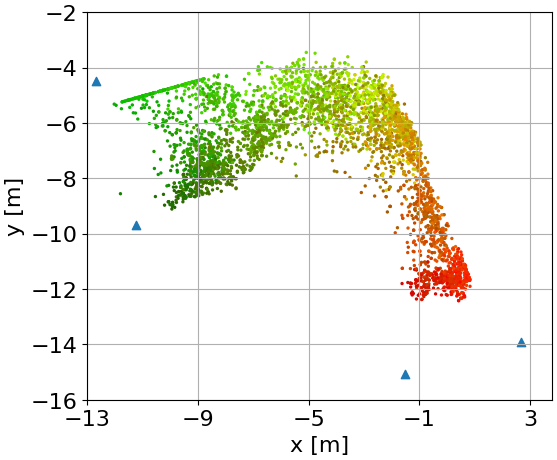}
    }}\hspace{-4pt}
    \subfigure[SCS-EE]  
    {
    \label{fig:dichasus_scs_ee_pos}
    \resizebox{0.18\textwidth}{3.35cm}{\includegraphics{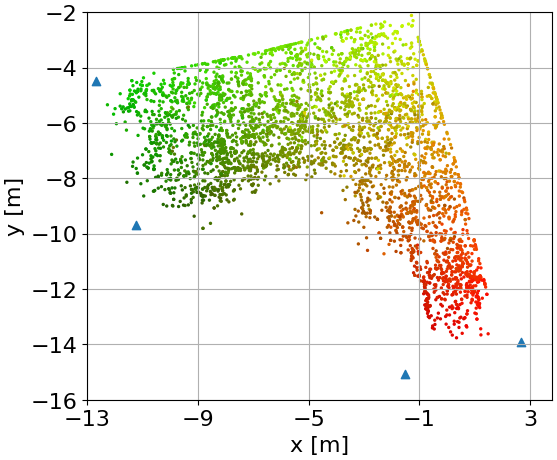}
    }}\hspace{-4pt}
    \subfigure[AE-AUG]
    {
    \label{fig:dichasus_ae_aug_pos}
    \resizebox{0.18\textwidth}{3.35cm}{\includegraphics{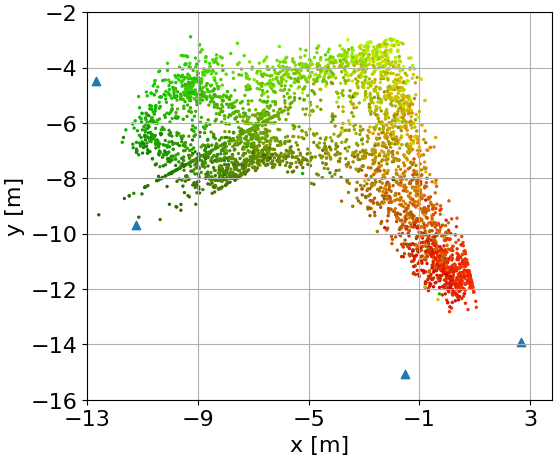}
    }}\hspace{-4pt}
     \subfigure[CSI2Vec] 
    {
    \label{fig:dichasus_csi2vec_cc}
    \resizebox{0.18\textwidth}{3.35cm}
    {\includegraphics{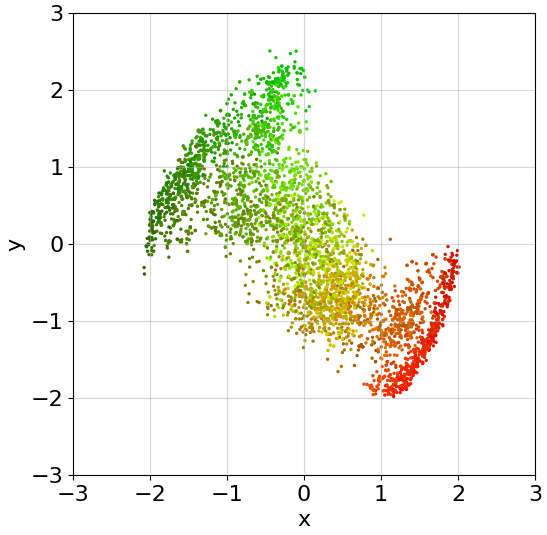}
    }}\hspace{-4pt}
    \subfigure[CSI2Vec-AUG]
    {
    \label{fig:dichasus_csi2vec_aug_cc}
    \resizebox{0.18\textwidth}{3.35cm}
    {\includegraphics{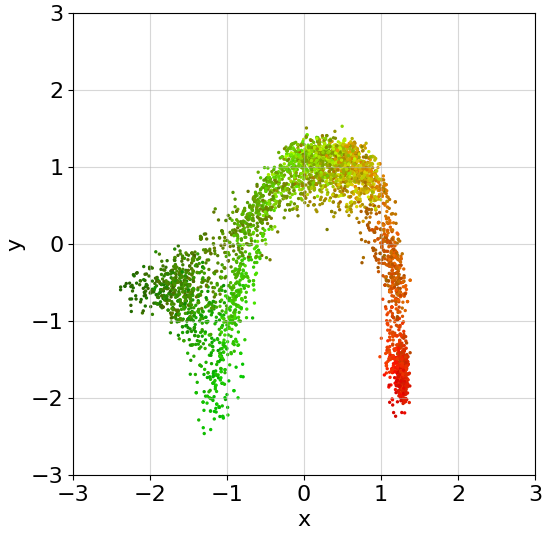}
    }}\hspace{-4pt}
    \subfigure[CSI2Vec-AUG-SEMI]
    {
    \label{fig:dichasus_csi2vec_aug_ss_cc}
    \resizebox{0.18\textwidth}{3.35cm}
    {\includegraphics{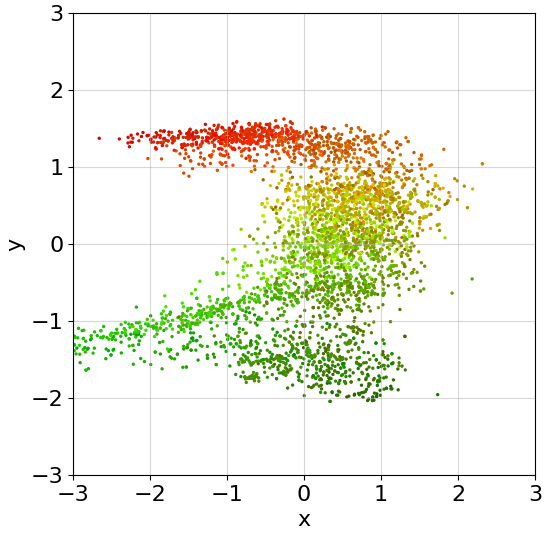}
    }}\hspace{-4pt}
    \subfigure[SCS-EE]  
    {
    \label{fig:dichasus_scs_ee_cc}
    \resizebox{0.18\textwidth}{3.35cm}
    {\includegraphics{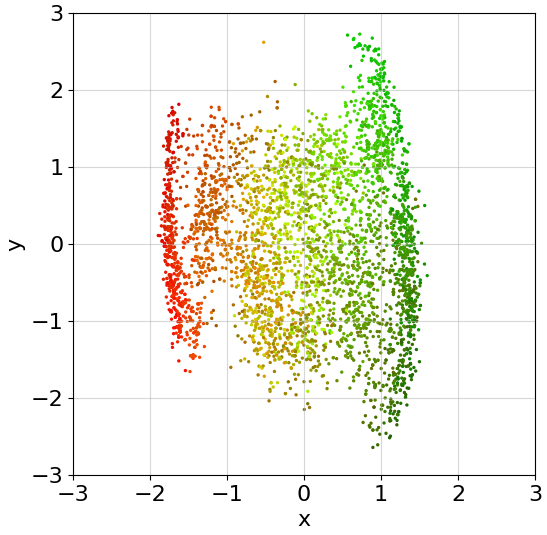}
    }}\hspace{-4pt}
    \subfigure[AE-AUG]
    {
    \label{fig:dichasus_ae_aug_cc}
    \resizebox{0.18\textwidth}{3.35cm}
    {\includegraphics{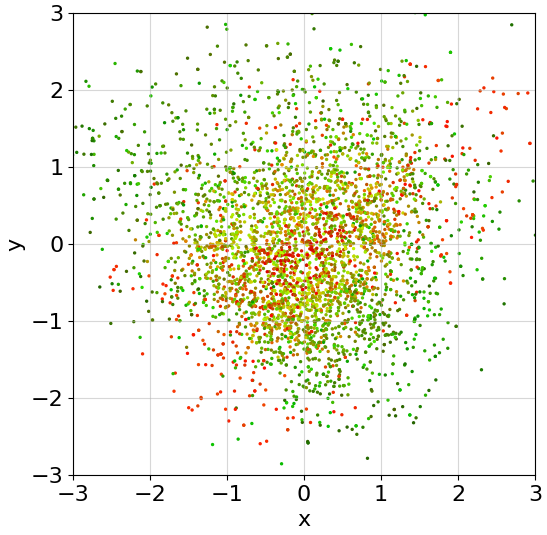}
    }}\hspace{-4pt}
    \caption{Results for the measurement-based indoor scenario: (a-e) POS estimates and (f-j) channel charts using CC-SN for both the proposed and baseline methods. 
    The proposed CSI2Vec-AUG method achieves an MDE, which is only $0.566$\,m higher than that of the SCS-EE baseline for POS, and delivers comparable CC performance to SCS-EE for CC-SN, despite using a $64\times$ smaller input feature dimension. 
    The AE-AUG baseline succeeds in producing a coherent estimated POS map but fails to produce a coherent channel chart.
    } 
    \label{fig:dichasus_results}  
\end{figure*}

In Figs. \ref{fig:dichasus_csi2vec_pos}-(e), we present the estimated POS maps for both the proposed methods and baselines. 
As observed, all methods produce well-shaped POS maps that preserve the color gradient and closely resemble the ground-truth UE positions. 
\fref{tbl:measured_indoor} shows that CSI2Vec-AUG achieves an {MDE which is only $0.566$\,m higher than that of the SCS-EE baseline}, despite using significantly smaller input features (i.e., only $D'=16$ instead of  $1024$).
Additionally, AUG improves POS performance, whereas CSI2Vec with SEMI does not provide an advantage, as the ground-truth UE positions do not belong to the evaluated scenario. Examining the estimated POS map of AE-AUG, we observe that in this scenario, having ground-truth UE positions for all samples is sufficient to establish spatial relationships between samples, leading to good POS performance. 
We reach the same conclusion for the~AE.
 
In Figs. \ref{fig:dichasus_csi2vec_cc}-(j) and \fref{tbl:measured_indoor}, we present the channel charts for CC-SN and the performance metrics for CC-SN and CC-PCA, respectively, for both the proposed methods and baselines. 
As observed, all methods---except for AE-AUG---produce channel charts that preserve the color gradient of the ground-truth UE positions. 
The results indicate that the CSI2Vec methods generate channel charts comparable in quality to those obtained through SCS-EE, despite using significantly smaller input features (i.e., only $D'=16$ instead of  $1024$).
Examining the performance metrics of AE-AUG and AE, we observe that their vector embeddings fail entirely to capture spatial relationships, resulting in poor channel charts.

\begin{table*}
\centering
\hspace{-0.25cm}
\resizebox{0.99\textwidth}{!}{
\begin{minipage}[c]{0.99 \textwidth}
    \centering
    \caption{POS and CC performance comparison for the measurement-based indoor scenario.} 
    \label{tbl:measured_indoor}
    \begin{tabular}{@{}lccccccccc@{}}
        \toprule
        & & & & \multicolumn{2}{c}{Positioning error [m]} & \multicolumn{4}{c}{Latent space quality metrics} \\
        \cmidrule(lr){5-6} \cmidrule(lr){7-10}  
        Method & Downstream task & Figure & Dim. & Mean$\,\downarrow$ & 95th percentile$\,\downarrow$ & TW$\,\uparrow$ & CT$\,\uparrow$ & KS$\,\downarrow$ & RD$\,\downarrow$ \\
        \midrule
        \multirow{3}{*}{CSI2Vec (across scenarios)} 
        & POS & \fref{fig:dichasus_csi2vec_pos} & 16 & 1.244 & 2.700 & 0.915 & 0.933 & 0.226 & 0.832 \\ 
        & CC-SN & \fref{fig:dichasus_csi2vec_cc} & 16 & -- & -- & 0.902 & 0.922 & 0.301 & 0.889 \\
        & CC-PCA & -- & 16 & -- & -- & 0.903 & 0.923 & 0.300 & 0.890\\
        \midrule
        \multirow{3}{*}{CSI2Vec-AUG (across scenarios)} 
        & POS & \fref{fig:dichasus_csi2vec_aug_pos} & 16 & 1.191 & 2.703 & 0.913 & 0.932 & 0.222 & 0.830 \\ 
        & CC-SN & \fref{fig:dichasus_csi2vec_aug_cc} & 16 & -- & -- & 0.896 & 0.920 & 0.309 & 0.895 \\
        & CC-PCA & -- & 16 & -- & -- & 0.885 & 0.914 & 0.300 & 0.892\\
        \midrule
        \multirow{3}{*}{CSI2Vec-AUG-SEMI (across scenarios)} 
        & POS & \fref{fig:dichasus_csi2vec_aug_ss_pos} & 16 & 1.218 & 2.709 & 0.912 & 0.929 & 0.225 & 0.836  \\ 
        & CC-SN & \fref{fig:dichasus_csi2vec_aug_ss_cc} & 16 & -- & -- & 0.893 & 0.915 & 0.320 & 0.899\\
        & CC-PCA & -- & 16 & -- & -- & 0.875 & 0.907 & 0.319 & 0.895\\
        \midrule
        \multirow{3}{*}{SCS-EE} 
        & POS & \fref{fig:dichasus_scs_ee_pos} & 1024 & 0.625 & 1.904 & 0.966 & 0.968 & 0.138 & 0.712 \\ 
        & CC-SN & \fref{fig:dichasus_scs_ee_cc} & 1024 & -- & -- & 0.905 & 0.901 & 0.289 & 0.896 \\
        & CC-PCA & -- & 1024 & -- & -- & 0.910 & 0.926 & 0.275 & 0.879 \\
        \midrule
        \multirow{3}{*}{AE (across scenarios)} 
        & POS & -- & 16 & 1.349 & 3.061 & 0.889 & 0.896 & 0.258 & 0.865 \\ 
        & CC-SN & -- & 16 & -- & -- & 0.684 & 0.581 & 0.613 & 0.986 \\
        & CC-PCA & -- & 16 & -- & -- & 0.663 & 0.571 & 0.747 & 0.993 \\
        \midrule
        \multirow{3}{*}{AE-AUG (across scenarios)} 
        & POS & \fref{fig:dichasus_ae_aug_pos} & 16 & 1.102 & 2.479 & 0.927 & 0.933 & 0.222 & 0.836 \\ 
        & CC-SN & \fref{fig:dichasus_ae_aug_cc} & 16 & -- & -- & 0.600 & 0.549 & 0.608 & 0.997 \\
        & CC-PCA & -- & 16 & -- & -- & 0.568 & 0.534 & 0.646 & 0.999\\
        \bottomrule
    \end{tabular}		
\end{minipage}}
\end{table*}

\subsection{Main Takeaways}
\label{sec:takeaways}

From the previously shown results, we reach the following main conclusions: 
\begin{itemize}
    \item For POS, CSI2Vec-based methods generally achieve a slightly higher MDE than the baseline SCS-EE. 
    However, CSI2Vec generates \emph{universal} vector embeddings, i.e., trained across all scenarios and with embedding dimension $64\times$ smaller than the original input features. 
    \item For CSI2Vec followed by downstream tasks, the overall number of activations of the neural networks from input to output is $\{1024,32,16,12,8,6,4,2\}$, whereas for SCS-EE, the overall number of activations is $\{1024,320,160,80,40,20,10,5,2\}$, clearly showing the complexity and storage advantage of our proposed method for inference.
    \item For POS, CC-SN, and CC-PCA, CSI2Vec-based methods with AUG not only generate more accurate POS maps and channel charts but also take into account variations in the channel vectors, such as different RSs.
    \item CSI2Vec with SEMI provides only a slight improvement, and only in scenarios where ground-truth UE position information is available.
    This observation suggests that self-supervised CSI2Vec is sufficient for most cases.
    \item AEs, with and without AUG, trained across scenarios, fail to produce useful POS maps except for one of the scenarios (measurement-based indoor scenario). 
    This indicates that their vector embeddings lack the universality of CSI2Vec.
    \item In CC-SN and CC-PCA downstream tasks, AEs with and without AUG fail to produce meaningful channel charts. 
    This suggests that their vector embeddings do not preserve spatial relationships, making them unsuitable for POS and CC downstream tasks.
    \item {The limitations of AE-based architectures in performing POS and CC tasks can be attributed to two main factors: 
    (i) AEs do not inherently preserve the spatial relationships between CSI samples, and (ii) as highlighted in \cite{Studer2018}, POS and CC do not operate directly on raw CSI (i.e., with small-scale fading), which is crucial for effective CSI reconstruction. 
    A potential approach to enable POS and CC with AE-based architectures is presented in \cite{Huang2019}, where the authors introduce pairwise vector embedding constraints into the AE framework. 
    However, even in this method, CSI compression is applied to the CSI features (cf. \fref{sec:feature_extraction}) rather than to the raw CSI data.}
\end{itemize}

\subsection{Impact of Vector Embedding Dimension on Performance}

\begin{figure*}[h]
    \centering
    \subfigure[Simulated outdoor scenario]
    {
    \includegraphics[width=0.30\textwidth,height=4cm]{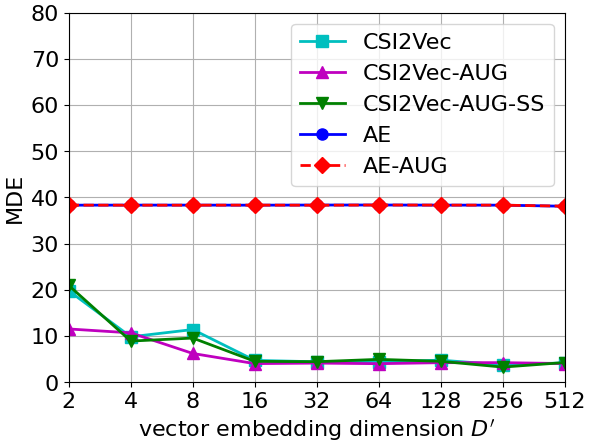}
    }
    \subfigure[Simulated indoor scenario]  
    {
    \includegraphics[width=0.30\textwidth,height=4cm]{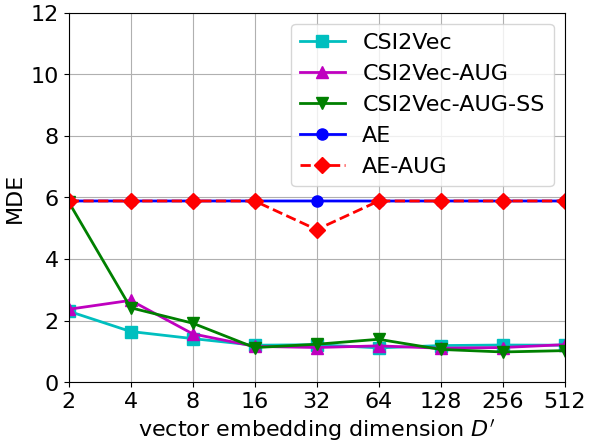}
    }
    \subfigure[Measurement-based indoor scenario]
    {
    \includegraphics[width=0.30\textwidth,height=4cm]{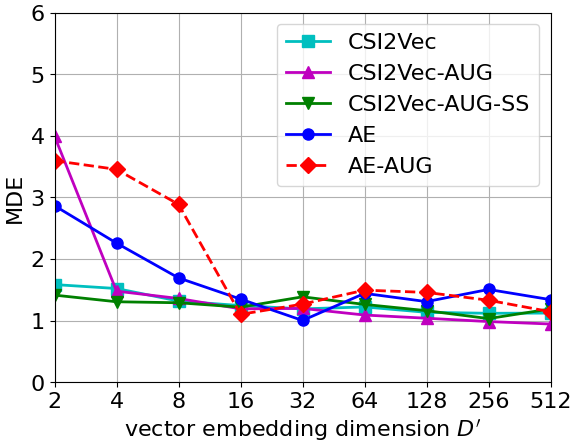}
    }
    \caption{MDE of both the proposed and baseline methods when varying the vector embedding dimension $D'$.
    For all the proposed CSI2Vec methods, MDE decreases as $D'$ increases, reaching the best performance at $D'=16$. 
    For AE-AUG and AE baselines, MDE remains consistently high, except for the measurement-based indoor scenario, where the AE-based approaches follow the same trend as the proposed CSI2Vec methods.} 
    \label{fig:vector_embedding_dim}  
\end{figure*}

We now evaluate the impact of the dimension of the vector embeddings---produced by the CSI2Vec function $g_{\boldsymbol{\phi}} \PC{\cdot}$ and the encoding function $g_{\boldsymbol{\varphi}} \PC{\cdot}$---on POS performance for all scenarios. 
This study aims to determine (i) the most suitable vector embedding dimension for all methods and (ii) whether the AE-AUG and AE baselines can produce meaningful vector embeddings for any given dimension $D'$, since they failed at $D'=16$ in all but one scenario (the measurement-based indoor scenario). 
\fref{fig:vector_embedding_dim} presents the result for the POS downstream task. 
We observe that for all proposed CSI2Vec-based methods, the MDE decreases as the vector embedding dimension increases, reaching the best performance at $D'=16$. 
Meanwhile, the AE-AUG and AE baselines completely fail to produce useful POS maps (up to an embedding dimension of $512$) except for the measurement-based indoor scenario, where they follow the same trend as CSI2Vec (i.e., MDE decreases as the vector embedding dimension increases, providing the best performance at $D'=16$).
We conclude that (i) $D'=16$ is an excellent choice for the CSI2Vec 
embedding dimension, striking a balance between POS accuracy and low MLP complexity, and (ii) the persistently high MDE for AE-based methods in the simulated outdoor and indoor scenarios highlights their lack of universality, even with higher-dimensional vector embeddings.


\section{Conclusions and Future Work}
\label{sec:conclusions}
We have proposed CSI2Vec, a self-supervised framework for generating universal CSI vector embeddings for positioning and channel charting. 
CSI2Vec is not only robust across diverse deployment and radio setups but also effectively encodes spatial information into compact representations, enabling hardware abstraction and reducing the computational complexity of both data transmission and processing. In addition, the computed vector embeddings can hide possible hardware specifics that manufacturers might not want to release. 

Our experiments demonstrate that CSI2Vec captures spatial relationships across three different scenarios, achieving a vector embedding $64\times$ smaller than the original feature size while only slightly degrading positioning accuracy and preserving channel charting performance. 
Compared to autoencoder-based methods, CSI2Vec was proven to be better suited for tasks that depend on spatial relationships.

Several promising directions remain for future work. 
First, applying CSI2Vec to many more scenarios and datasets is left for future work. 
Second, enforcing similarity constraints across scenarios might further improve the universality of the proposed CSI2Vec method.
Third, incorporating new forms of data augmentation, such as selectively using certain user equipment antennas or accounting for other hardware impairments, might further enhance the robustness of our method.

\bibliographystyle{IEEEtran}

\balance

\bibliography{bib/IEEEabrv,bib/confs-jrnls,bib/publishers,bib/studer, bib/CSI2Vec_bib} 

\balance 
	
\end{document}